\documentclass[pra, reprint]{revtex4-2}
\usepackage{bm}
\usepackage{graphicx}
\usepackage{siunitx}
\usepackage[final]{microtype}
\usepackage{amsmath, amsfonts, amssymb}
\usepackage{xcolor}
\usepackage{braket}
\usepackage[normalem]{ulem}
\usepackage{soul}
\usepackage{afterpage}

\DeclareSIUnit\wn{\raiseto{-1}\cm}

\begin{document}

\title{Model for two-body collisions between ultracold dipolar molecules \\
around a F{\"o}rster resonance in an electric field}

\author{Lucas Lassabli{\`e}re, Goulven Qu{\'e}m{\'e}ner}
\affiliation{Universit\'{e} Paris-Saclay, CNRS, Laboratoire Aim\'{e} Cotton, 91405, Orsay, France}

\begin{abstract}
We propose a one-channel, simple model to describe the dynamics of ultracold dipolar molecules
around a F{\"o}rster resonance. Slightly above a specific electric field, a collisional shielding can take place, suppressing the molecular losses in a gas.
The overall description of the quantum physical mechanism 
comes back to the dynamics on a unique energy surface, 
which depends on the relative distance and angular approach of the molecules.
This surface enables to interpret how the dipole moments of the molecules are induced
and interlocked by the electric field  and the dipole-dipole interaction during the process, 
especially when the shielding is triggered.
Averaging the relative angular motion over a unique partial wave 
(the lowest one when the ultracold regime is reached), the model
reproduces well the behaviour of the rate coefficients 
observed experimentally and predicted theoretically 
[Matsuda et al., Science 370, 1324 (2020); Li et al., Nat. Phys. 17, 1144 (2021)].
This economic model encapsulates the main physics of the quantum process. Therefore,
it can be used as an alternative to a full quantum dynamical treatment
and is promising for future studies of collisions involving more bodies.
\end{abstract}
\maketitle

\section{Introduction}

In ultracold molecular gases, two-body loss collisions strongly limit the lifetime of molecules. The loss processes dominate over the elastic ones at the typical temperatures of hundreds of nanokelvins reached in the experiments, leading to unfavorable conditions for long-lived gases,  
efficient evaporative cooling and quantum degeneracy. The overall losses, often referred to as quenching processes, include either reactive processes when chemical reactions are possible \cite{Ospelkaus_S_327_853_2010, Ni_N_464_1324_2010} or complex formation processes 
\cite{Takekoshi_PRL_113_205301_2014,Molony_PRL_113_255301_2014,Park_PRL_114_205302_2015,
Guo_PRL_116_205303_2016,Ye_SA_4_eaaq0083_2018,Guo_PRX_8_041044_2018,Gregory_NC_10_3104_2019,
Voges_PRL_125_083401_2020,Bause_PRR_3_033013_2021}.
If the molecules are initially prepared in excited rovibrational states, 
inelastic processes can also occur and contribute to the quenching.  
All these collisional processes, which can happen either for ultracold 
bosons or for fermions, are then highly problematic and 
prevent the exploration of 
many interesting applications~\cite{Carr_NJP_11_055049_2009,Bohn_S_357_1002_2017,Schmidt_PRR_4_013235_2022}.
It is therefore very important to find ways to shield and protect 
the molecules against these bad collisions.

For that purpose, different methods have been theoretically and experimentally investigated. They exploit the long-range properties of the dipole-dipole interaction to generate a repulsive barrier in the lowest partial wave of the incident channel ($s$-wave for indistinguishable bosons, $p$-wave for indistinguishable fermions).
If the barrier is high enough compared to the initial collision energy, the molecules remain at long-range and the quenching rate is suppressed. 
A first solution can use a dc electric field in a confined geometry to force the molecules to start to collide in a side-by-side configuration
so that the dipole-dipole interaction is repulsive~
\cite{Ticknor_PRA_81_042708_2010,Quemener_PRA_81_060701_2010,Micheli_PRL_105_073202_2010,
DeMiranda_NP_7_502_2011,Quemener_PRA_83_012705_2011,Simoni_NJP_17_013020_2015,
Frisch_PRL_115_203201_2015}.
A second solution consists in using dipolar molecules initially prepared in their first excited rotational 
state~\cite{Avdeenkov_PRA_73_022707_2006,Wang_NJP_17_035015_2015,Gonzalez-Martinez_PRA_96_032718_2017,Matsuda_S_370_1324_2020,Li_NP_17_1144_2021}.
At a particular electric field, one can bring a couple of rotational states of a two-body collision in resonance with the initial one so that their energies are degenerate. 
This process is reminiscent of a 
F{\"o}rster resonance~\cite{Forster_AP_437_55_1948,Walker_JPBAMOP_38_S309_2005,
Walker_PRA_77_032723_2008,Comparat_JOSAB_27_208_2010}.
If the initial couple of rotational states is slightly above the other one, then the strong dipole-dipole couplings between those quantum states create a barrier at long range while generating moderate inelastic collisions, preventing losses to occur for the initial molecules. 
A third solution consists in shielding molecules initially prepared in their ground rovibrational states by appropriately dressing them with a 
microwave field~\cite{Micheli_PRA_76_043604_2007,Gorshkov_PRL_101_073201_2008,
Lassabliere_PRL_121_163402_2018,
Karman_PRL_121_163401_2018,Karman_PRA_100_052704_2019,Karman_PRA_101_042702_2020,
Anderegg_S_373_779_2021,Schindewolf_arXiv_2201_05143_2022}.
This method provides additional tools to control the stability of such gases through the control of the sign and the value of the molecular scattering length~\cite{Lassabliere_PRL_121_163402_2018,Schmidt_PRR_4_013235_2022}. 
A last method is based on the use of an optical field that couples different electronic states, 
proposed initially for two atoms~\cite{Napolitano_PRA_55_1191_1997} and extended for 
two molecules~\cite{Xie_PRL_125_153202_2020}.

In this paper, we focus on the second solution, motivated by recent experimental 
observations of the static electric field shielding on fermionic 
$^{40}$K$^{87}$Rb molecules in a one dimensional optical lattice 
in a quasi two-dimensional geometry~\cite{Matsuda_S_370_1324_2020}
and then in a free-space collisional environment without the presence
of a confining optical lattice~\cite{Li_NP_17_1144_2021}. 
In these two experiments, suppression/enhancement 
of two-body quenching losses was observed
around a F{\"o}rster resonance in an electric field, 
in quantitative agreement with previous theoretical predictions
based on a time-independent quantum formalism~\cite{Wang_NJP_17_035015_2015}.
Then, the loss suppression seen in the free space, 
non-confined environment, led to efficient evaporative cooling of the molecules
\cite{Li_NP_17_1144_2021,Wang_PRA_103_063320_2021}.
Here, we pursue our theoretical investigation
and present a simple model which simplifies the two-body collisional
problem at its best, sufficient to semi-quantitatively describes
the physical collisional process.
This model extracts and preserves the essential physics 
of the two-body resonant mechanism involved in the
electric field shielding method, 
requiring only a small amount of numerical effort. 
This model can represent an important basis to start 
extended investigations of few- and many-body interactions/collisions
involving dipolar molecules in electric fields.

The paper is organized as follows. In Sec.~\ref{section:model}, we present 
the model based on four simplifications. 
The first one consists in using only zero angular projection quantum numbers to describe the rotational structure of the molecules as well as the orbital motion.
The second one consists in removing
explicitly the rotational structure out of the problem and replacing it by the values of the induced dipole moments of the molecules as well as their transition dipole moments, in an electric field. 
The third one will focus on the two most important combined molecular states 
for the dynamics (initial and resonant state).
The fourth one imposes that the collisional problem is described
by a unique partial wave, the lowest one.
These four assumptions set the basis of the model
and greatly simplify the collisional formalism.
In Sec.~\ref{section:rates}, we compare the rate coefficients obtained with the model with a quantum calculation for both bosonic and fermionic KRb molecules.
Finally,  we conclude in Sec.~\ref{section:conclusion}.

\section{A simple, effective collisional model}
\label{section:model}

Our aim is to provide a simple, effective model for the collision
of two ultracold molecules around a F{\"o}rster resonance, 
removing unnecessary considerations and parameters 
while keeping the essential physics untouched, 
in order to reduce the numerical calculation at its maximum.
\\

Generally in nowadays experiments, all the molecules
of an ultracold gas can be formed in an arbitrary 
electronic, vibrational and rotational quantum state.
In the electric field shielding experiments of fermionic $^{40}$K$^{87}$Rb molecules 
that we are interested to describe~\cite{Matsuda_S_370_1324_2020,Li_NP_17_1144_2021},
the molecules are prepared in the electronic and vibrational ground state, but in the 
first excited rotational state 
$\big| \tilde{j}, m_{j} \big\rangle = \big| \tilde{1}, 0 \big\rangle$.
Here $\big| {j}, m_{j} \big\rangle$ is the rotational quantum number of the molecule.
The tilde notation $\big| \tilde{j}, m_{j} \big\rangle$ 
characterizes the notation of the rotational state in the electric field.
This ``dressed'' rotational state
is a linear combination of several rotational states 
with a main character in $\big| {j}, m_{j} \big\rangle$. Namely for a given 
electric field ${\cal E}$ and $m_j$ component, we have
\begin{eqnarray}\label{dressedrot}
\big| \tilde{j}, m_{j} \big\rangle = \sum_{j} \big| {j}, m_{j} \big\rangle \, 
\langle {j}, m_{j} \big| \tilde{j}, m_{j} \big\rangle.
\end{eqnarray}
To describe two-body collisions, one needs to 
define combined molecular states
$\ket{\tilde{j}_1,m_{j_1}}\ket{\tilde{j}_2,m_{j_2}}$
where $j_\tau, m_{j_\tau}$ are the rotational 
quantum numbers of molecules $\tau=1,2$. 
More precisely, as the molecules are identical, we define 
properly symmetrized combined molecular states
\begin{multline}\label{symCMS}
\left\{ \big| \tilde{j}_1, m_{j_1} \big\rangle  \, \big| \tilde{j}_2, m_{j_2} \big\rangle \right\}_\pm = 
\frac{1}{\sqrt{2(1+\delta_{\tilde{j}_1,\tilde{j}_2} \delta_{m_{{j}_1},m_{{j}_2}})}} \\
\bigg\{ \big| \tilde{j}_1, m_{j_1} \big\rangle \, \big| \tilde{j}_2, m_{j_2}  \big\rangle \pm \big| \tilde{j}_2, m_{j_2} \big\rangle \, \big| \tilde{j}_1, m_{j_1}  \big\rangle \bigg\}.
\end{multline}
In the following, only symmetric symmetrized states involving a $+$ sign in the previous equation will be needed.
We also omit here the electronic and vibrational degrees of freedom 
as they are not involved in such experimental conditions.
In the experiment, the electric field is tuned in such a way that
the energies of two combined molecular state become 
degenerate~\cite{Avdeenkov_PRA_73_022707_2006,Wang_NJP_17_035015_2015}.
This happens at a resonant electric field ${\cal E}^* \simeq 12.5$~kV/cm 
(for fermionic $^{40}$K$^{87}$Rb).
This defines the condition 
of a F{\"o}rster resonance~\cite{Forster_AP_437_55_1948,Walker_JPBAMOP_38_S309_2005,
Walker_PRA_77_032723_2008,Comparat_JOSAB_27_208_2010}.
The two combined molecular states of interest are 
the initial one, noted $\big| 1 \big\rangle$, with 
\begin{eqnarray}\label{symCMSinit}
\big| 1 \big\rangle \equiv \left\{ \big| \tilde{1}, 0 \big\rangle  \, \big| \tilde{1}, 0 \big\rangle \right\}_+
\end{eqnarray}
of energy $E_{| 1 \rangle} = E_{| \tilde{1}, 0 \rangle} +
E_{| \tilde{1}, 0 \rangle}$,
and the resonant one, noted $\big| 2 \big\rangle$, with
\begin{eqnarray}\label{symCMSreso}
\big| 2 \big\rangle \equiv \left\{ \big| \tilde{0}, 0 \big\rangle  \, \big| \tilde{2}, 0 \big\rangle \right\}_+ 
\end{eqnarray}
of energy $E_{| 2 \rangle} = E_{| \tilde{0}, 0 \rangle} +
E_{| \tilde{2}, 0 \rangle}$.
These energies depend on the electric field ${\cal E}$, but the explicit dependence is omited.
\\

\begin{figure*}[t!]
\begin{center}
\includegraphics*[width=8cm, trim=0cm 0cm 0cm 0cm]{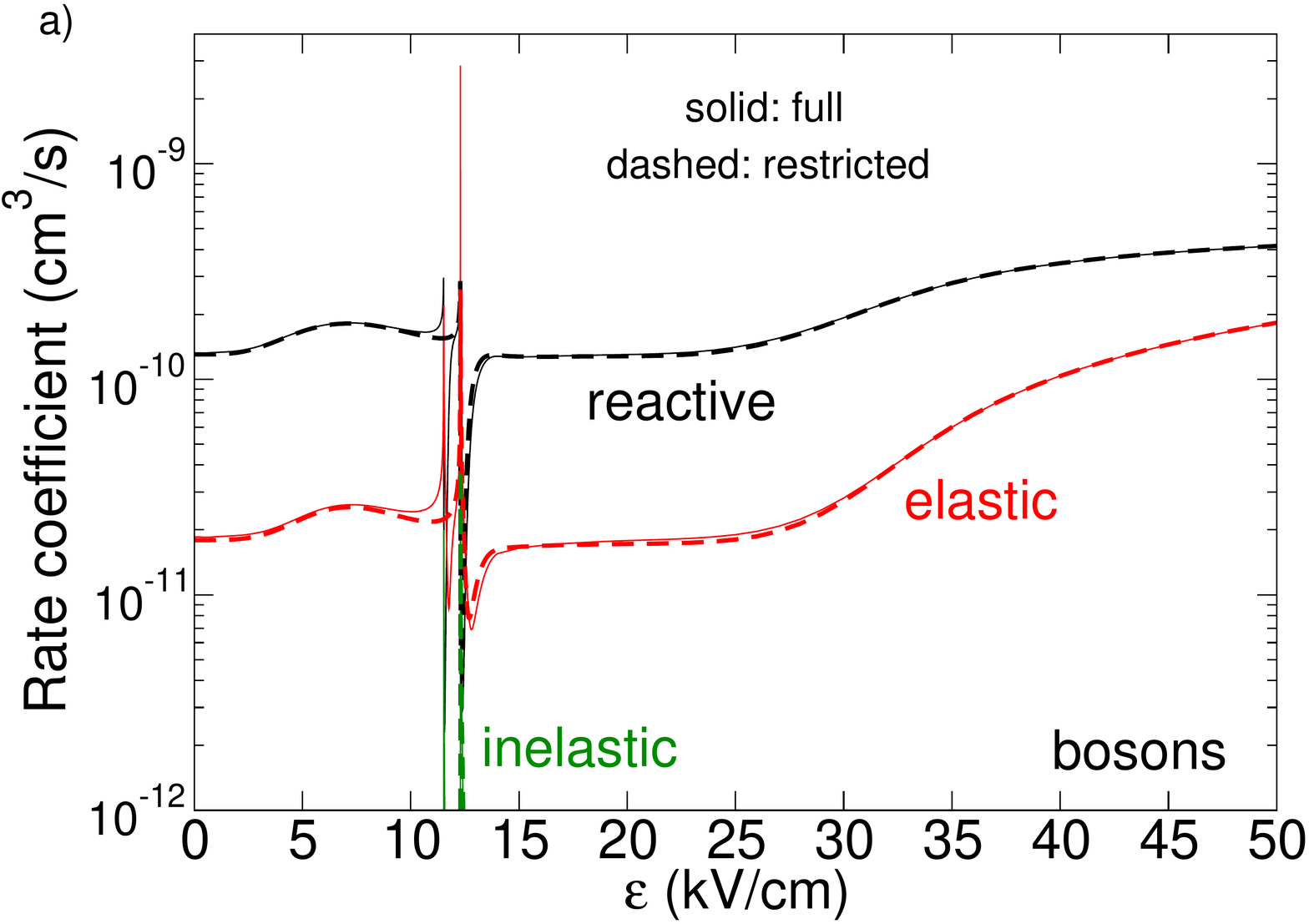} 
\includegraphics*[width=8cm, trim=0cm 0cm 0cm 0cm]{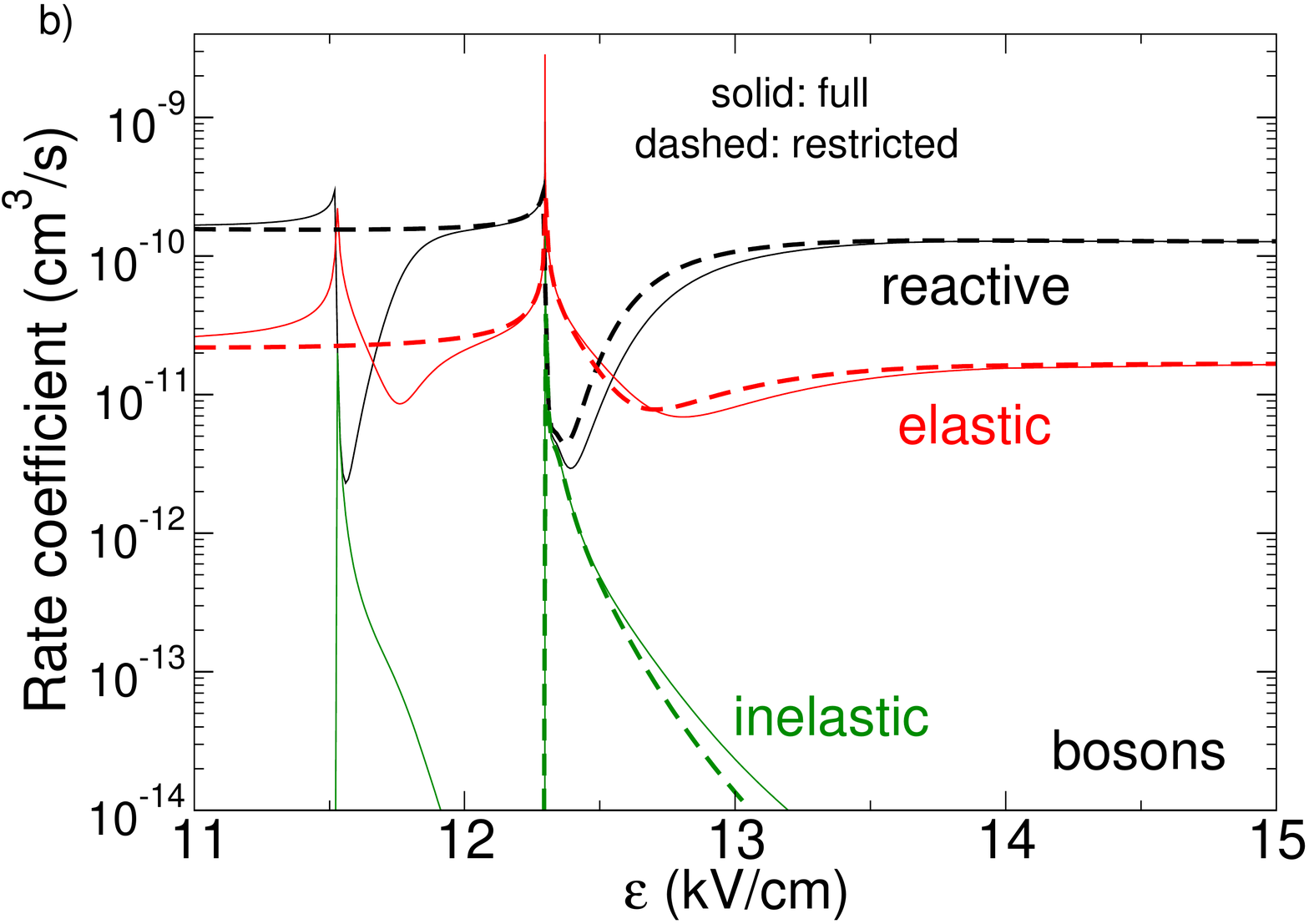} \\
\includegraphics*[width=8cm, trim=0cm 0cm 0cm 0cm]{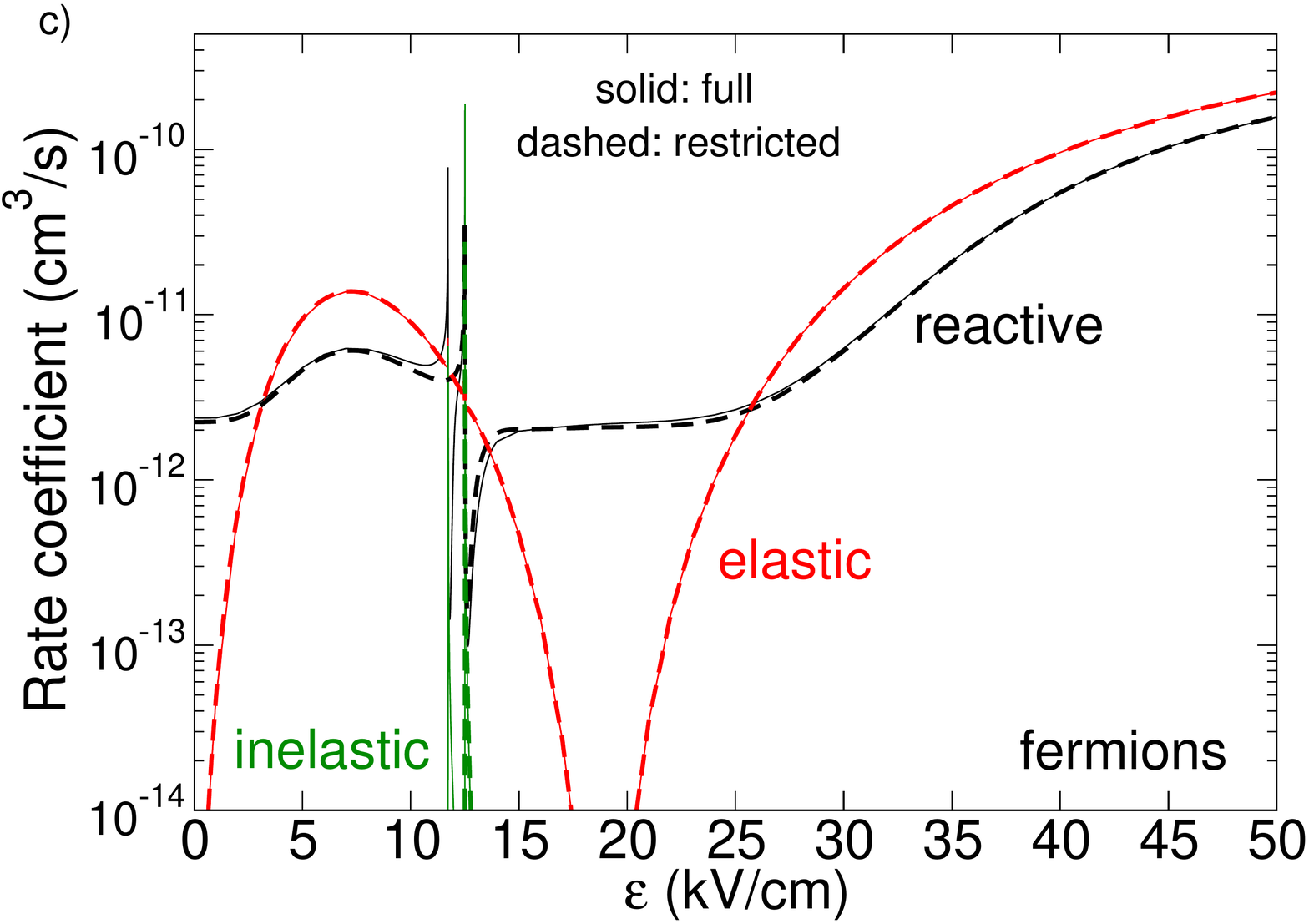} 
\includegraphics*[width=8cm, trim=0cm 0cm 0cm 0cm]{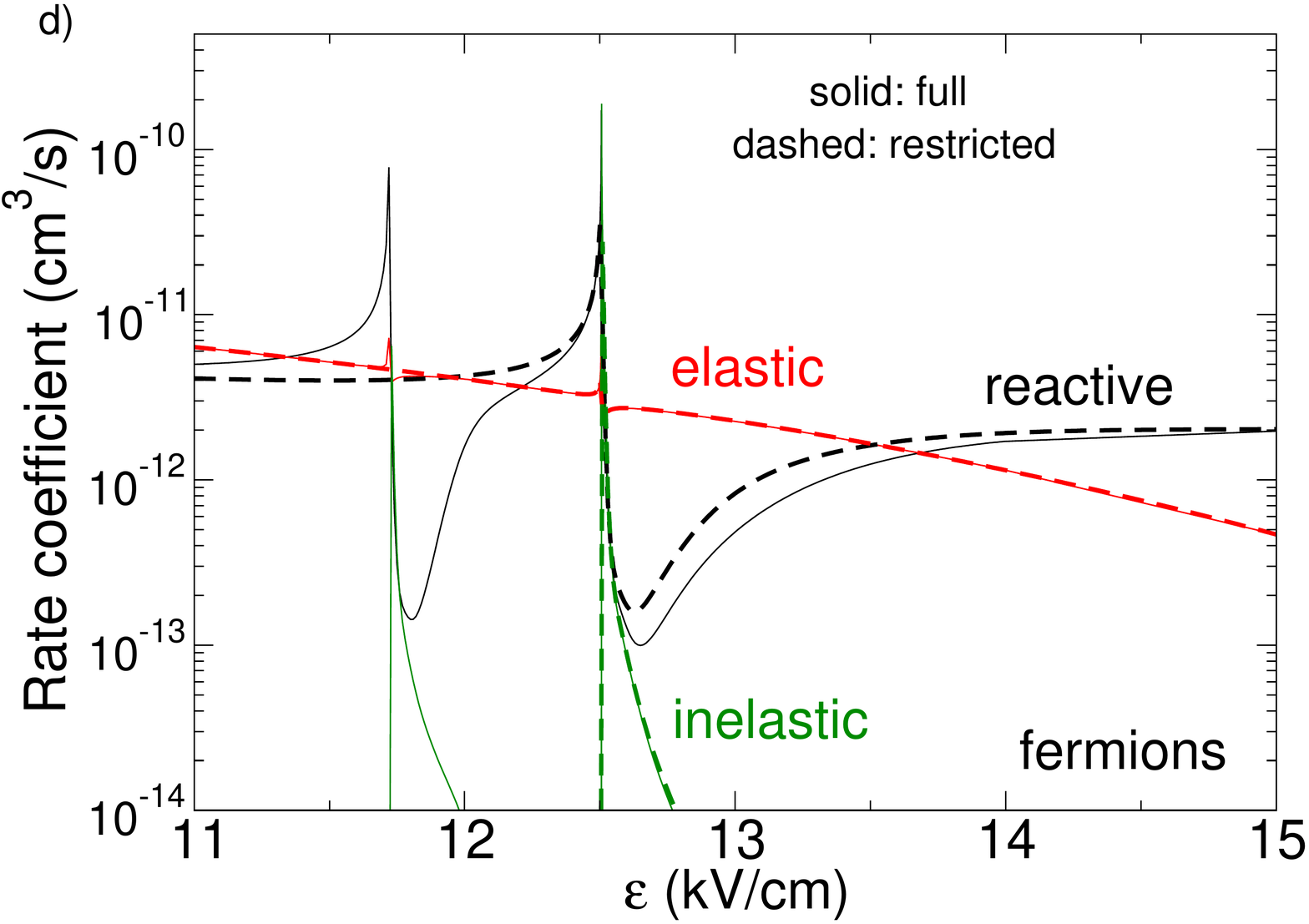}
\caption{Rate coefficients as a function of the electric field
for bosonic $^{41}$K$^{87}$Rb (panels a, b) and fermionic $^{40}$K$^{87}$Rb (panels c, d) for a collision energy $E_c=\SI{500}{\nano\kelvin}$. The b and d panels correspond to zooms of the panels a and c.
% $E_{\text{k}}=\SI{500}{\nano\kelvin}$. 
Black line: reactive process, red line: elastic process, green line: inelastic process. Solid lines: a full-quantum calculation with $0 \le j_\tau \le 5$ with all allowed $m_{j_\tau}$, for molecules $\tau=1,2$.
Values of $l=0,2,4$, $M = 0$ are taken 
for bosons and $l=1,3,5$, $M = 0,\pm1$ for fermions. 
Dashed lines: a restricted calculation with in addition all $m_{j_\tau}=0$.}
\label{FIG-FULL}
\end{center}
\end{figure*}

Then, to describe the total wavefunction of the colliding system 
of the two molecules,
we use an usual partial wave expansion to complete 
the internal basis set in Eq.~\eqref{symCMS}
(see~\cite{Wang_NJP_17_035015_2015} for more details). 
The basis set used is then
\begin{eqnarray}
\left\{ \big| \tilde{j}_1, m_{j_1} \big\rangle  \, \big| \tilde{j}_2, m_{j_2} \big\rangle \right\}_\pm \, \big| l, m_l \big\rangle
\end{eqnarray}
where in this study we took
$0 \le j_\tau \le 5$.
For indistinguishable bosons, we took $l=0,2,4$ 
and a total projection quantum number 
$M=m_{j_1}+m_{j_2}+m_l=0$.
For indistinguishable fermions, we took $l=1,3,5$ 
and $M=0,\pm1$.
From Eq.~\eqref{dressedrot} 
and Eq.~\eqref{symCMS}, this basis set can be expressed 
in term of the basis set
$\big| {j}_1, m_{j_1} \big\rangle  \, \big| {j}_2, m_{j_2} \big\rangle \, \big| l, m_l \big\rangle$,
in which the dipole-dipole interaction is given by
\begin{multline}\label{Vdd-full}
\left\langle j_{1}, m_{j_1}, j_{2}, m_{j_2}, l, m_{l} \left| V_{dd} \right| j_{1}^{\prime}, m_{j_1}^{\prime}, j_{2}^{\prime}, m_{j_2}^{\prime}, l^{\prime}, m_{l}^{\prime} \right\rangle 
= \\
- \frac{2 d_{1} d_{2}}{4 \pi \epsilon_{0} r^{3}}(-1)^{m_{l}} \\ 
\times \sqrt{\left(2 j_{1}+1\right)\left(2 j_{1}^{\prime}+1\right)}
\left(\begin{array}{ccc}
j_{1} & 1 & j_{1}^{\prime} \\
0 & 0 & 0
\end{array}\right) 
\left(\begin{array}{ccc}
j_{1} & 1 & j_{1}^{\prime} \\
-m_{j_1} & 0 & m_{j_1}^{\prime} 
\end{array}\right) \\
\times \sqrt{\left(2 j_{2}+1\right)\left(2 j_{2}^{\prime}+1\right)}
\left(\begin{array}{ccc}
j_{2} & 1 & j_{2}^{\prime} \\
0 & 0 & 0
\end{array}\right)
\left(\begin{array}{ccc}
j_{2} & 1 & j_{2}^{\prime} \\
-m_{j_2} & 0 & m_{j_2}^{\prime} 
\end{array}\right) \\
\times\sqrt{(2 l+1)\left(2 l^{\prime}+1\right)}  
\left(\begin{array}{ccc}
l & 2 & l^{\prime} \\
0 & 0 & 0
\end{array}\right)\left(\begin{array}{ccc}
l & 2 & l^{\prime} \\
-m_{l} & 0 & m_{l}^{\prime}
\end{array}\right). 
\end{multline}
The quantities $d_1, d_2$ correspond to the permanent electric dipole moments
of molecules 1, 2.
In this basis set, we consider the electronic van der Waals interaction 
to be diagonal, as treated in previous studies (see for example 
~\cite{Wang_NJP_17_035015_2015}), so that
\begin{multline}\label{VvdW}
\left\langle j_{1}, m_{j_1}, j_{2}, m_{j_2}, l, m_{l} \left| V_{vdW} \right| j_{1}^{\prime}, m_{j_1}^{\prime}, j_{2}^{\prime}, m_{j_2}^{\prime}, l^{\prime}, m_{l}^{\prime} \right\rangle \\
= - \frac{C_{6}^{e l}}{r^6} \times \delta_{j_{1}, {j_1}'}  \, \delta_{m_{j_{1}}, m_{{j_1}'}}  
\, \delta_{j_{2}, {j_2}'}  \, \delta_{m_{j_{2}}, m_{{j_2}'}}  
\, \delta_{l,l'}  \, \delta_{m_l,m_l'}. 
\end{multline}
We use the experimental values of the rotational constant $B=\SI{1.113950}{\GHz}$~\cite{Ospelkaus_PRL_104_030402_2010} for 
$^{40}\mathrm{K}^{87} \mathrm{Rb}$ 
and $B=\SI{1.095362}{\GHz}$~\cite{Aikawa_PRL_105_203001_2010}
for $^{41}\mathrm{K}^{87} \mathrm{Rb}$. 
We use the experimental value of the permanent 
dipole moment $d=0.574$ D~\cite{Ni_PhDThesis_2009} 
and the theoretical electronic van der Waals coefficient 
$C_{6}^{e l}=12636$ a.u~\cite{Lepers_PRA_88_032709_2013,Vexiau_JCP_142_214303_2015}.

\subsection{First simplification: Setting $m_{j_1} = m_{j_2} = 0$}
\label{subsection:mj_approximation}

In previous studies~\cite{Quemener_PRA_93_012704_2016,Gonzalez-Martinez_PRA_96_032718_2017}, it was shown that 
rotational states, for which only $m_{j_1} = m_{j_2} = 0$ 
in the full-quantum calculation and in Eq.~\eqref{Vdd-full} were used,
provided a good approximation around the electric field
of the F{\"o}rster resonance. 
We then follow this simplification in our model.
\\

We plot in Fig.~\ref{FIG-FULL} as solid lines, 
the elastic (red), inelastic (green) and reactive (black) rate coefficients
using the full-quantum calculation employed 
in~\cite{Wang_NJP_17_035015_2015,Quemener_PRA_93_012704_2016,Gonzalez-Martinez_PRA_96_032718_2017} (referred to as full calculation in the following). 
Fig.~\ref{FIG-FULL}-a and Fig.~\ref{FIG-FULL}-b corresponds to collisional properties
of bosonic $^{41}\mathrm{K}^{87} \mathrm{Rb}$ molecules with ${\cal E}^* \simeq 12.2$~kV/cm,
while Fig.~\ref{FIG-FULL}-c and Fig.~\ref{FIG-FULL}-d corresponds to fermionic 
$^{40}\mathrm{K}^{87} \mathrm{Rb}$ molecules with ${\cal E}^* \simeq 12.5$~kV/cm.
Note that for fermions, the full calculation was in very good agreement with the 
experimental results of~\cite{Matsuda_S_370_1324_2020,Li_NP_17_1144_2021}.
To validate the aforementioned simplification,
we also plot as dashed lines the rate coefficients
when the combined molecular states are restricted 
to $m_{j_1} = m_{j_2} = 0$ in the quantum calculation 
(referred to as restricted calculation in the following). 
\\

Over a large range of electric fields (panels a and c), 
the full and restricted calculations are globally the same for all processes.
It means that it is sufficient to restrict values of $m_{j_\tau} = 0$
to describe the overall background process, 
away from the F{\"o}rster resonances.
Looking over a closer range of electric fields (panels b, d) where the resonances
are present, one can see some differences.
First, the resonances that appear around a field of ${\cal E} \simeq 11.5$~kV/cm 
and ${\cal E} \simeq 11.7$~kV/cm for the bosonic and fermionic cases 
in the full treatment (solid lines),
are absent in the restricted treatment (dashed lines).
This is expected as this F{\"o}rster resonance involves the initial state
$\left\{ \big| \tilde{1}, 0 \big\rangle  \, \big| \tilde{1}, 0 \big\rangle \right\}_+$
but namely the resonant state
$\left\{ \big| \tilde{0}, 0 \big\rangle  \, \big| \tilde{2}, \pm 1 \big\rangle \right\}_+ $.
As the restricted calculation keeps only values of $m_{j_\tau} = 0$,
it cannot obviously treat states with $m_{j_\tau} = \pm1$
so that the corresponding resonance are missing. 
However and to be compliant with the aforementionned experiment, 
we are interested here in the second resonances 
that appear around a field of ${\cal E} \simeq 12.2$~kV/cm and ${\cal E} \simeq 12.5$~kV/cm
for the bosonic and fermionic cases,
where the initial state is still 
$\left\{ \big| \tilde{1}, 0 \big\rangle  \, \big| \tilde{1}, 0 \big\rangle \right\}_+$
but the corresponding resonant state is
$\left\{ \big| \tilde{0}, 0 \big\rangle  \, \big| \tilde{2}, 0 \big\rangle \right\}_+ $.
This state is treated in the restricted calculation
and now these resonances are present
and well taken into consideration within the first simplification.
The elastic and reactive processes are globally well described by the dashed lines,
meaning that the entrance collisional channel is slightly affected by the restriction 
employed and the removal of all states containing $m_{j_\tau} \ne 0$. This is in agreement
with the previous studies~\cite{Quemener_PRA_93_012704_2016,Gonzalez-Martinez_PRA_96_032718_2017} where a comparison between 
the entrance collisional channels with or without restrictions was quite 
good at ultralow energies.
While the difference of the rate coefficients
can reach about a factor of two, 
the trend is respected, keeping the physical mechanism untouched.
The inelastic process is also well described by the dashed lines, 
as the resonant state responsible
for the inelastic transition from
$\left\{ \big| \tilde{1}, 0 \big\rangle  \, \big| \tilde{1}, 0 \big\rangle \right\}_+$
to
$\left\{ \big| \tilde{0}, 0 \big\rangle  \, \big| \tilde{2}, 0 \big\rangle \right\}_+ $
has $m_{j_\tau} = 0$ components and well taken into account.

\subsection{Second simplification: Removal of the rotational structure}

We now go further in the simplification by removing explicitly the 
rotational quantum numbers out of the problem.
We implicitly consider the use of the first simplification $m_{j_1} = m_{j_2} = 0$ 
and we omit their notation in the following, so that the two combined molecular states of interest in Eq.~\eqref{symCMSinit} and Eq.~\eqref{symCMSreso} become
$\big| 1 \big\rangle \equiv \left\{ \big| \tilde{1} \big\rangle  \, \big| \tilde{1} \big\rangle \right\}_+$
and
$\big| 2 \big\rangle \equiv \left\{ \big| \tilde{0} \big\rangle  \, \big| \tilde{2} \big\rangle \right\}_+$.
To simplify the formalism, we employ the definition of a generalized induced dipole moment between a state $\tilde{j}$ and $\tilde{j}'$ at a given electric field ${\cal E}$
\begin{multline}\label{induced_dipole_2B}
d^{\tilde{j} \to \tilde{j}'} = d  \, \sum_{j ,j'} 
\langle \tilde{j} \, | \, {j}  \rangle  \,
\langle {j'} \, | \, \tilde{j}' \rangle \\ 
\sqrt{2 j+1} \, \sqrt{2 j'+1} \,
\left( \begin{array}{ccc} j  & 1 & j' \\ 0 & 0 & 0 \end{array} \right)^2.
\end{multline}
The derivation is given in appendix~\ref{app_GIDM}.
This corresponds to the generalization of the 
expectation value of the permanent dipole moment 
of a molecule in a specific electric field between two states $\tilde{j}$ and $\tilde{j}'$,
when a dipole-dipole interaction takes place with another molecule. 
When $\tilde{j} = \tilde{j}'$, the
generalized induced dipole moment identifies directly with the induced dipole moment.
When $\tilde{j} \ne \tilde{j}'$,  the generalized induced dipole moment 
can be seen as a transition dipole moment from $\tilde{j}$ $\to$ $\tilde{j}'$. 
This definition was also used 
to characterize the dynamics of interacting dipolar molecules
siting at fixed positions of a lattice \cite{Gorshkov_PRA_84_033619_2011,Gorshkov_PRL_107_115301_2011}.
The quantities $\langle {j} \big| \tilde{j} \big\rangle$ and $\langle {j'} \big| \tilde{j'} \big\rangle$
are the coefficients of the development of the dressed states as 
expressed in Eq.~\eqref{dressedrot}, omitting the $m_j$ notations.
Using Eq.~\eqref{induced_dipole_2B}, the expression of the dipole-dipole interaction
in Eq.~\eqref{Vdd-full} in the dressed state basis Eq.~\eqref{dressedrot}
simplifies to an effective classical expression
\begin{eqnarray}\label{Vdd-classic}
V_{dd}^{\tilde{j}_1 \to \tilde{j}_1', \tilde{j}_2 \to \tilde{j}_2'}(r,\theta) = \frac{d^{\tilde{j}_1 \to \tilde{j}_1'} \, d^{\tilde{j}_2 \to \tilde{j}_2'}}{4 \pi \varepsilon_0 \, r^3} \, (1 - 3 \cos^2 \theta)
\end{eqnarray}
similar to the dipole-dipole interaction between two classical dipoles, 
but now where a general (elastic or inelastic) 
transition $(\tilde{j}_1, \tilde{j}_2) \to (\tilde{j}_1', \tilde{j}_2')$ is taken into account
in an effective way. \\

\begin{figure*}[t]
\begin{center}
\includegraphics*[width=7cm, trim=0cm 10cm 12cm 0cm]{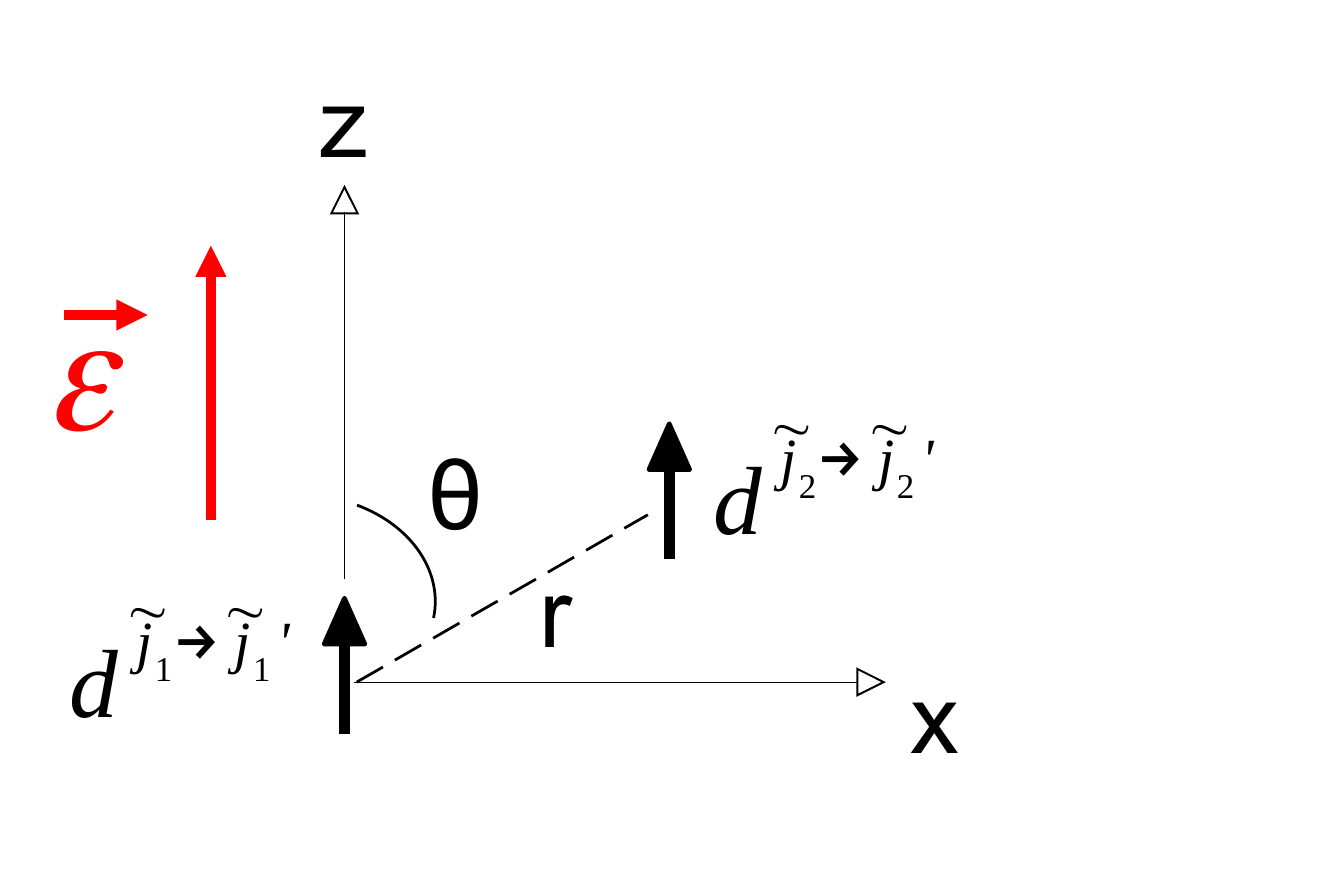} 
\includegraphics*[width=5.2cm, trim=0cm 7cm 12cm 0cm]{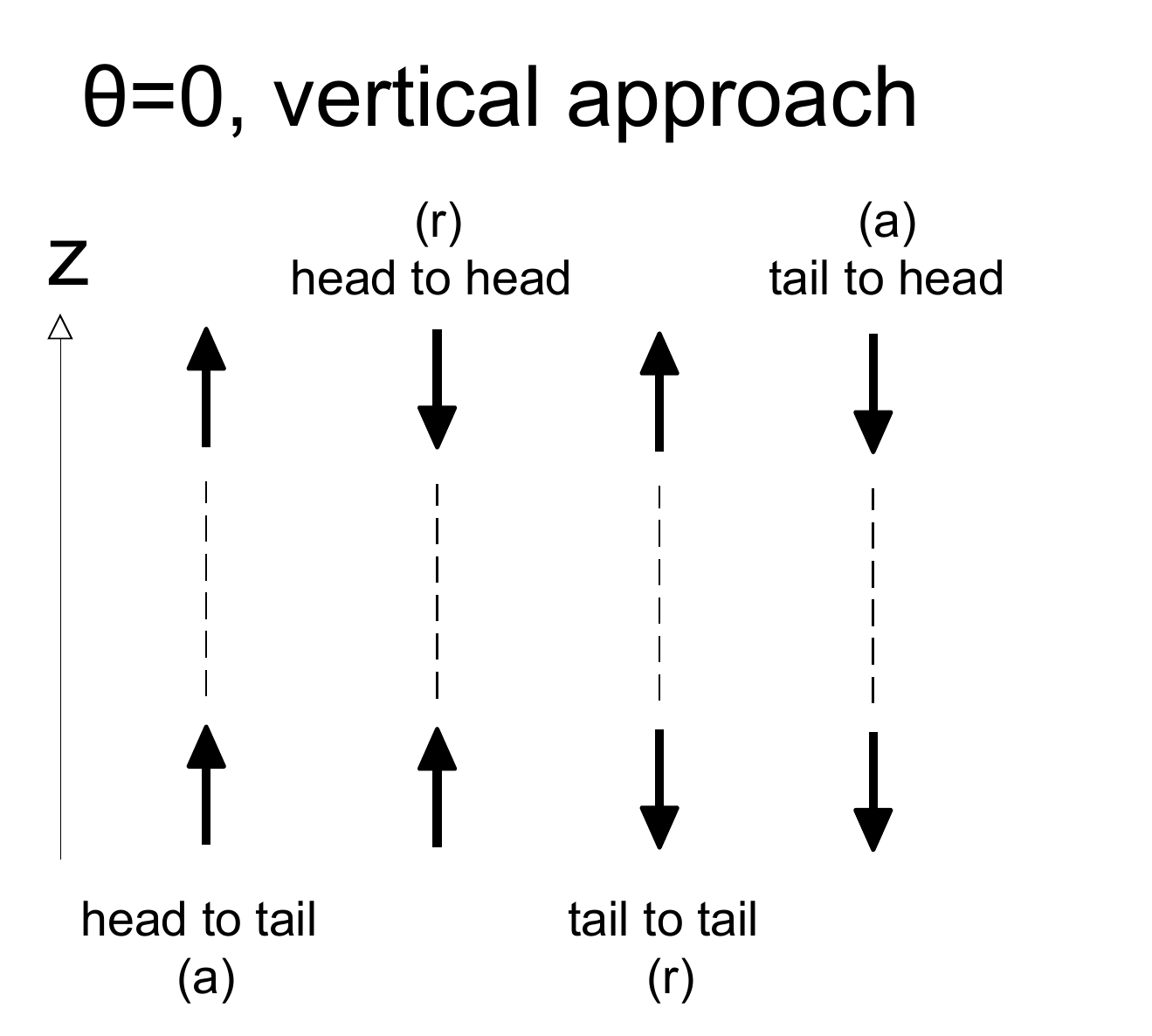} 
\includegraphics*[width=5.5cm, trim=0cm 7cm 11cm 0cm]{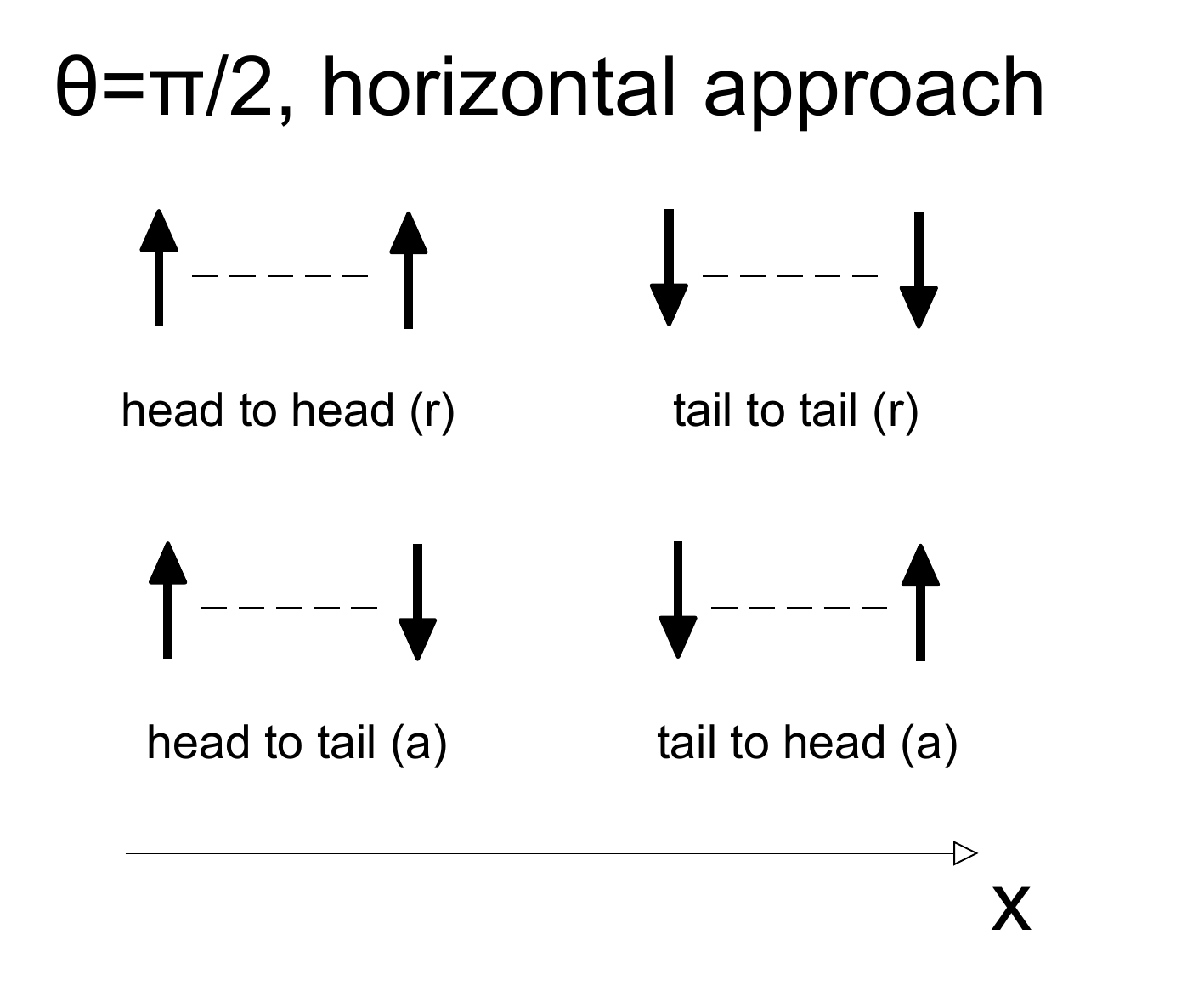}
\caption{Left: Sketch of a dipole-dipole interaction between two induced dipole moments $d^{\tilde{j}_1 \to \tilde{j}_1'}$ and $d^{\tilde{j}_2 \to \tilde{j}_2'}$, here with positive values along the $z$ axis. Middle: When $\theta=0$, the approach is vertical. Depending on the sign of the induced dipoles, different dipolar approaches are allowed leading to different namings and types of interaction: attractive (a) or repulsive (r). The interaction can be attractive or repulsive. Right: Same as the middle panel but with $\theta=\pi/2$ for a horizontal approach.}
\label{FIG-2BODY-CONFIGURATION}
\end{center}
\end{figure*}

We illustrate Eq.~\eqref{Vdd-classic} in Fig.~\ref{FIG-2BODY-CONFIGURATION} showing 
the dipole-dipole interaction between two induced dipole moments $d^{\tilde{j}_1 \to \tilde{j}_1'}$ and $d^{\tilde{j}_2 \to \tilde{j}_2'}$, 
here with dipoles pointing up, i.e. positive values along the $z$ axis.
The angle $\theta$ characterizes the two-body approach, for example vertical when $\theta=0$ (middle panel), horizontal when $\theta=\pi/2$ (right panel). 
The signs of the induced dipoles characterize different dipolar interactions, namely
head to tail, head to head, tail to tail and tail to head. They are illustrated in the figure, they depend on the two-body approach and the angle $\theta$, and they 
can lead to attractive or repulsive interactions depending on the cases.
In the remaining of the paper, we will refer to the different namings shown on this figure
for the vertical and horizontal approaches.
\\

\begin{figure}[h]
\begin{center}
\includegraphics*[width=8cm, trim=0cm 0cm 0cm 0cm]{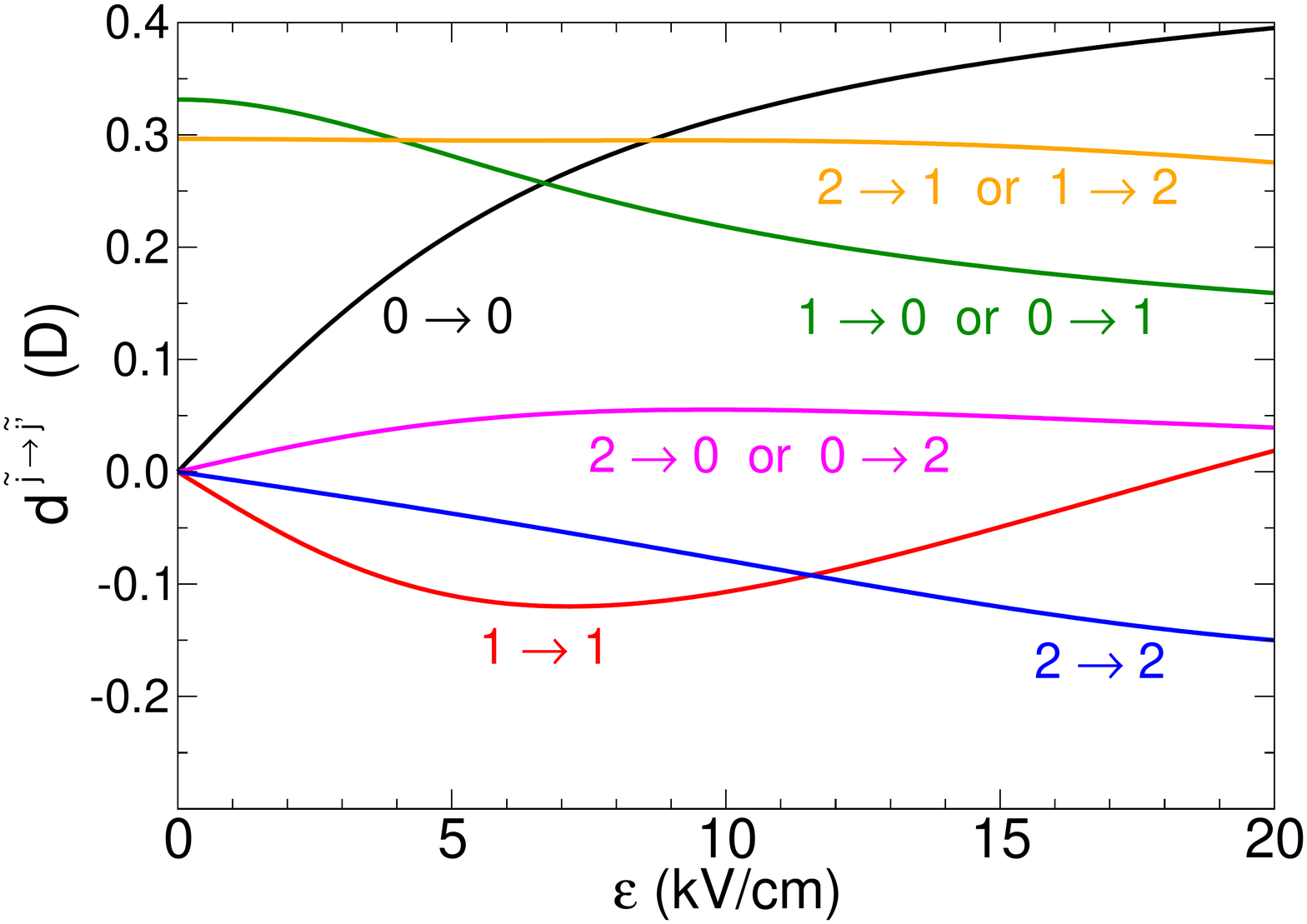}
\caption{Induced ($\tilde{j}' = \tilde{j}$) and transition ($\tilde{j}' \ne \tilde{j}$) dipole moments as a function of the electric field for different values $\tilde{j} \to \tilde{j}'$. We omitted the tilde in the dressed rotational state notations.}
\label{FIG-GIDMVSE}
\end{center}
\end{figure}

We present in Fig.~\ref{FIG-GIDMVSE} the induced and transition dipole moments as a function of the electric field ${\cal E}$ for different transitions $\tilde{j} = 0,1,2 \to \tilde{j}'=0,1,2$. 
The induced dipole moments involved in Eq.~\eqref{Vdd-classic}
in the elastic process of the 
initial combined molecular state 
$\big| 1 \big\rangle \to \big| 1 \big\rangle$ 
are $d^{1 \to 1}$ (red curve).
At the electric fields around the F{\"o}rster resonance,
namely ${\cal E}^* \simeq 12.2$~kV/cm for bosonic $^{41}\mathrm{K}^{87} \mathrm{Rb}$
and ${\cal E}^* \simeq 12.5$~kV/cm for fermionic 
$^{40}\mathrm{K}^{87} \mathrm{Rb}$, they are negative.
The orientations of both dipoles are then pointing 
against the electic field, say $\downarrow\downarrow$
(if we consider that the electric field is pointing up as 
in Fig.~\ref{FIG-2BODY-CONFIGURATION}).
Now, the induced dipole moments involved 
in the elastic process of the 
resonant state $\big| 2 \big\rangle $ $\to$ $\big| 2 \big\rangle $
can be separated in two, one for a direct process
and one for an exchange process due to the symmetrization of the states in Eq.~\eqref{symCMS}.
This can be seen later in the second line of 
Eq.~\eqref{2by2elmt}.
For the direct process, 
the induced dipole moments involved are $d^{0 \to 0}$ (black curve) and $d^{2 \to 2}$ (blue curve), which are positive and negative around ${\cal E}^*$, respectively.
The corresponding orientations are then pointing up and down, say $\uparrow\downarrow$.
For the exchange process, 
the induced dipole moments involved 
are $d^{0 \to 2} = d^{2 \to 0}$ (pink curve), 
which are positive around ${\cal E}^*$.
The corresponding orientations are then pointing up, say $\uparrow\uparrow$.
Finally, the transition dipole moments involved in the inelastic processes 
$\big| 1 \big\rangle $ $\leftrightarrow$ $\big| 2 \big\rangle $
are $d^{1 \to 0} = d^{0 \to 1}$ (green curve) 
and $d^{1 \to 2} = d^{2 \to 1}$ (orange curve),
which are all positive around ${\cal E}^*$.
\\

\subsection{Third simplification: Initial and resonant combined molecular states}

The third simplification will consist in using, via Eq.~\eqref{Vdd-classic}, 
only the initial and the resonant combined molecular states $\big| 1 \big\rangle $ 
and $\big| 2 \big\rangle$. 
This is somewhat dictated by the fact that around the F{\"o}rster resonance,
these two states are the main involved in the physical mechanism.
Then, the formalism is reduced to a two-level problem, involving a two-by-two matrix
\begin{eqnarray}\label{2by2mat}
\left[ \begin{array}{cc} 
E_a(r,\theta) & W(r,\theta) \\ 
W(r,\theta) & E_b(r,\theta)
\end{array}\right] 
\end{eqnarray}
with
\begin{eqnarray}\label{2by2elmt}
E_a(r,\theta) &=& \frac{d^{\tilde{1} \to \tilde{1}} \, d^{\tilde{1} \to \tilde{1}}}{4 \pi \varepsilon_0 \, r^3 } \, (1 - 3 \cos^2 \theta)+ E_{|1\rangle} \nonumber \\
E_b(r,\theta) &=& \frac{d^{\tilde{0} \to \tilde{0}} \, d^{\tilde{2} \to \tilde{2}}+ d^{\tilde{0} \to \tilde{2}} \, d^{\tilde{2} \to \tilde{0}}}{4 \pi \varepsilon_0 \, r^3} \, (1 - 3 \cos^2 \theta)+ E_{|2\rangle} \nonumber\\
W(r,\theta) &=& \sqrt{2} \, \frac{d^{\tilde{1} \to \tilde{0}} \, d^{\tilde{1} \to \tilde{2}}}{4 \pi \varepsilon_0 \, r^3} \, (1 - 3 \cos^2 \theta).
\end{eqnarray}
Note that in the second line of Eq.~\eqref{2by2elmt},
one can see clearly the direct and the exchange terms of Eq.~\eqref{Vdd-classic}
due to the symmetrization in Eq.~\eqref{symCMS}.
It is now useful to diagonalize the matrix in Eq.~\eqref{2by2mat}.
This results in eigenvalues given by two surface functions
in $r$ and $\theta$
\begin{eqnarray}\label{Epm}
E_{\pm}(r,\theta) = \frac{1}{2} (E_a + E_b) \pm \frac{1}{2} \sqrt{(E_a - E_b)^2 + 4 W^2}
\end{eqnarray}
associated with two eigenstates 
\begin{eqnarray}\label{ESpm}
\big| + \big\rangle &=& \ \  \alpha \, \big| {1} \big\rangle + \beta \, \big| {2} \big\rangle \nonumber \\
\big| - \big\rangle &=& -\beta \, \big| {1} \big\rangle + \alpha \, \big| {2} \big\rangle 
\end{eqnarray}
with
\begin{gather}
\alpha = \cos(\eta/{ 2}) \qquad \beta = \sin(\eta/{ 2})  \nonumber \\
\eta = \arctan \bigg[\frac{2 \, |W(r,\theta)|}{|E_a(r,\theta) - E_b(r,\theta)|} \bigg] .
\end{gather}
Each surface in Eq.~\eqref{Epm} defines an effective, 
global and anisotropic dipole-dipole interaction, 
which includes the effect of the dipolar couplings bewteen the combined molecular states. 
Hereafter, the energies $E_a, E_b$ will be referred to as diabatic energies (the coupling $W$
is not taken into account) and the quantities $E_+, E_-$ will be referred to as adiabatic
energies (where the coupling $W$ is taken into account).
For each given configuration in $r$ and $\theta$, one has to monitor what is the upper state
in energy and what is the lower state.
When $E_a(r,\theta) > E_b(r,\theta)$, the upper state with energy 
$E_a(r,\theta)$ takes the new value $E_+$ while the lower state with energy
$E_b(r,\theta)$ takes the new value $E_-$.
When $E_a(r,\theta) < E_b(r,\theta)$, the reverse is true.
Then at $r \to \infty$, if $E_{|1\rangle} > E_{|2\rangle}$, 
that is when the initial CMS lies above the resonant CMS 
(when ${\cal E} > {\cal E}^*$), 
this implies necessarily that the initial CMS is described by the surface $E_+$
and eigenstate $\big| + \big\rangle$. 
If $E_{|1\rangle} < E_{|2\rangle}$, 
that is when the initial CMS lies below the resonant CMS 
(when $E < E^*$),
this implies that the initial CMS is described by the surface $E_-$
and eigenstate $\big| - \big\rangle$. \\

\begin{figure}[t!]
\begin{center}
\includegraphics*[width=7.6cm, trim=0cm 0cm 0cm 0cm]{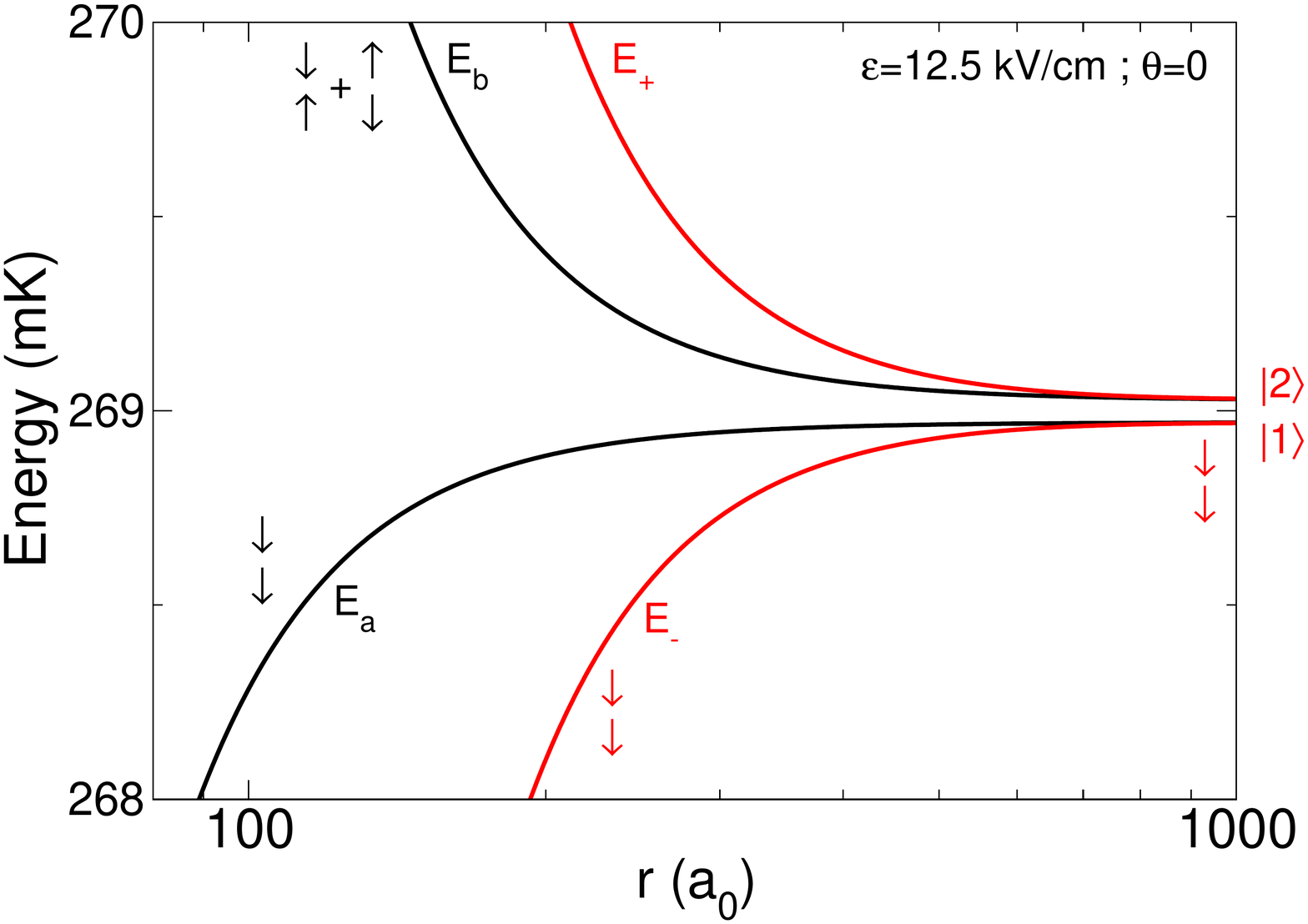} \\
\includegraphics*[width=7.6cm, trim=0cm 0cm 0cm 0cm]{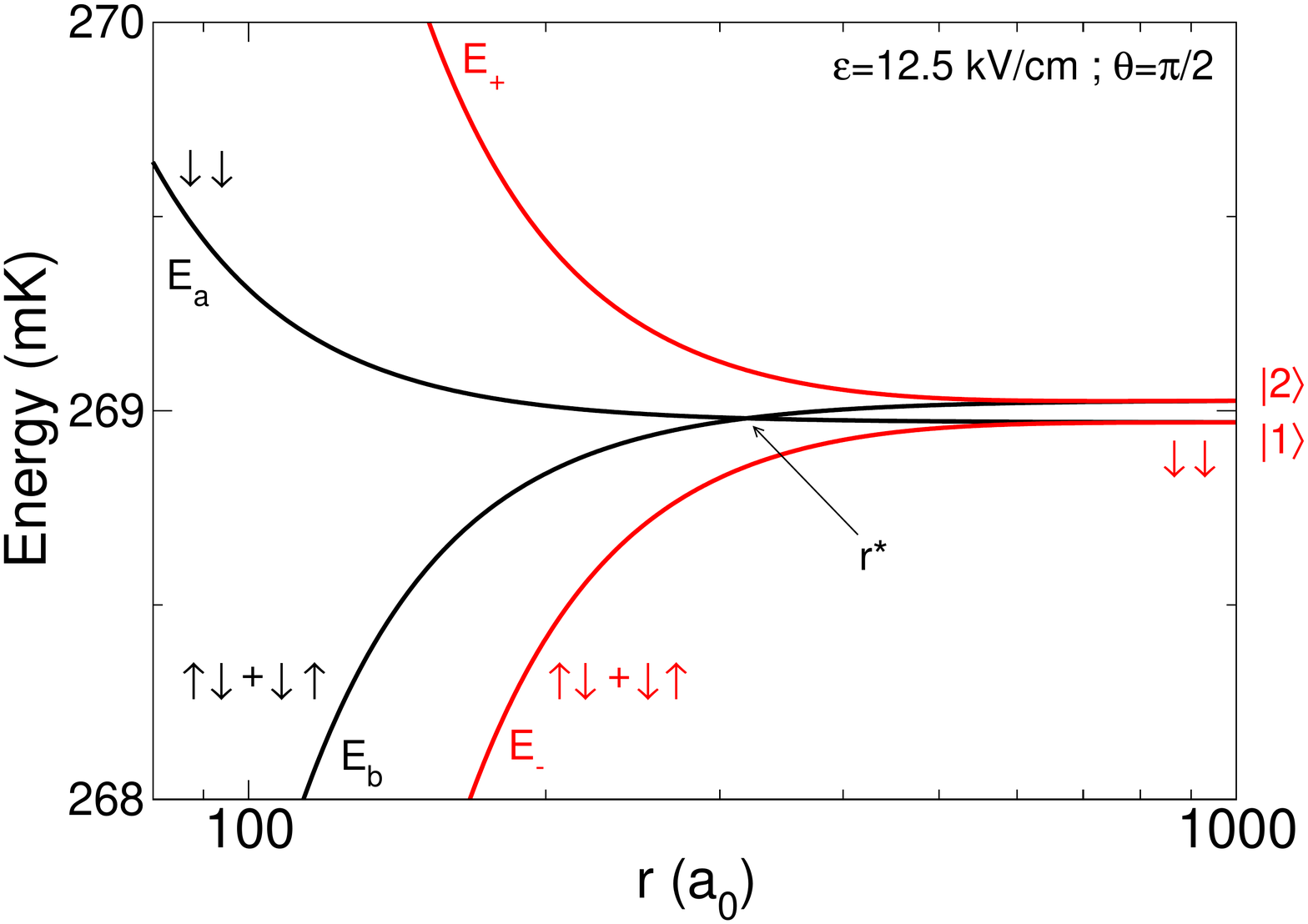} 
\caption{Diabatic $E_a$, $E_b$ (black curves) and 
adiabatic $E_{\pm}$ (red curves) energies
for $\theta=0$ (top) and $\theta=\pi/2$ (bottom) as a function of the intermolecular distance $r$ for ${\cal E}=12.5$~kV/cm. The diabatic energies cross at $r^*$.}
\label{FIG-EPM-1}
\end{center}
\end{figure}

\begin{figure}[t!]
\begin{center}
\includegraphics*[width=7.6cm, trim=0cm 0cm 0cm 0cm]{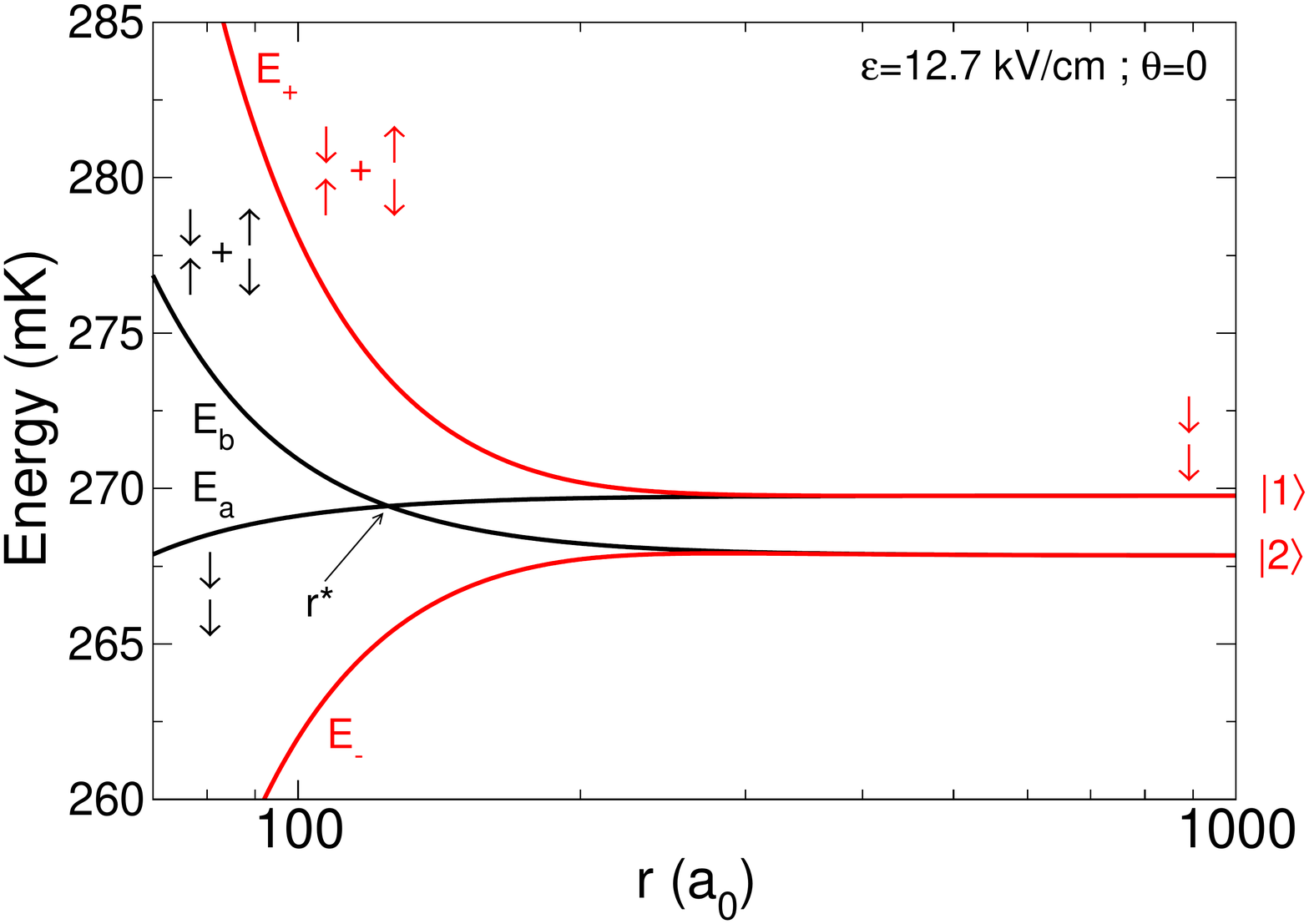} \\
\includegraphics*[width=7.6cm, trim=0cm 0cm 0cm 0cm]{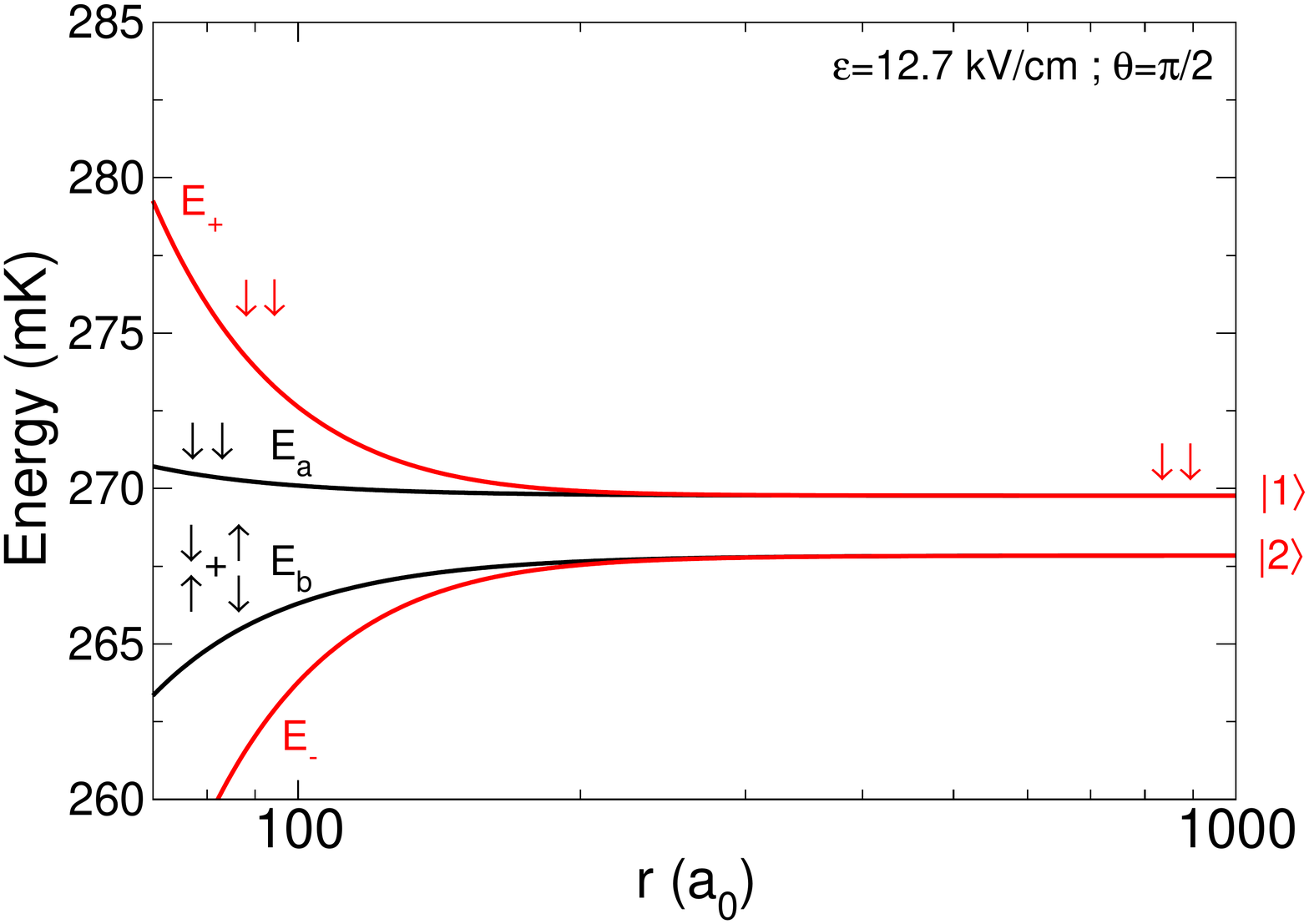} 
\caption{Same as Fig.~\ref{FIG-EPM-1} but for ${\cal E}=12.7$~kV/cm.}
\label{FIG-EPM-2}
\end{center}
\end{figure}

We assume within the model that the dynamics of the initial CMS 
will follow adiabatically the $E_+$ or $E_-$ surface, depending on the value of the electric field, and that it is sufficient to describe the main dynamics, at least at a semi-quantitative level. We will see in Section~\ref{section:rates} how good this approximation is.
This assumption is suggested by the ordering of the interactions strength.
The physical picture becomes the one of a dominant strong dipolar interaction first (this can persist at long-range as the two energy levels of the molecular states become more and more degenerate) due to the F{\"o}rster resonance, then of a subsequent coupling between partial waves. 
It is therefore physically intuitive to first look at the effect of the dipolar interaction (mediated by the rotational levels) which is the main contribution of the interaction, then to treat the effect on the orbital motion (mediated by the partial waves) 
at a second stage, as a pertubation. 
Note that this is in essence what was proposed and employed 
in~\cite{Gorshkov_PRL_101_073201_2008} 
where Born--Oppenheimer-like potentials were introduced to investigate similar studies of shielding with microwaves. 
\\

We illustrate the values of $E_+$ and $E_-$ 
in Fig.~\ref{FIG-EPM-1} for ${\cal E} = 12.5$~kV/cm
and Fig.~\ref{FIG-EPM-2} for ${\cal E} = 12.7$~kV/cm.
We plotted the diabatic energies $E_a, E_b$ (black curves) 
as well as the adiabatic ones $E_+, E_-$ (red curves), 
for two selected configurations of angles, 
$\theta=0$ (top panels) and $\theta=\pi/2$ (bottom panels).
For ${\cal E} = 12.5$~kV/cm, the energy of the initial CMS $\big| 1 \big\rangle$
is the lowest one as indicated in red on the figures at $r \to \infty$.
For $\theta=0$ in Eq.~\eqref{2by2elmt}
and because the product of the dipoles of the initial CMS
is positive (see Fig.~\ref{FIG-GIDMVSE}),
$E_a$ is attractive.
The approach is vertical tail to head (see black arrows).
In contrast, $E_b$ is a repulsive interaction because 
the overall term of the resonant CMS is negative
(namely the sum of the dipoles product from the direct and exchange terms 
in the second line of Eq.~\eqref{2by2elmt}).
This corresponds to the resonant CMS with a vertical head to head or tail to tail approach.
As $E_a$ remains smaller than $E_b$,
the initial CMS will follow adiabatically the curve $E_-$ from Eq.~\eqref{Epm},
and the molecules will collide with 
an overall vertical tail to head approach (see red arrows), corresponding to an overall attractive interaction.
But this can change with the angle.
For example for $\theta=\pi/2$ in Eq.~\eqref{2by2elmt}, 
$E_a$ is repulsive. The approach is horizontal tail to tail.
$E_b$ is now attractive corresponding to a horizontal head to tail or tail to head approach.
There is a certain distance, say $r^*$, where the two curves cross and where
$E_a$ becomes higher than $E_b$, so that the initial CMS
takes over the character of smallest energy, namely $E_b$. 
This is seen on the adiabatic energy curve $E_-$: while the molecules start to collide 
with a tail to head approach, they
will change of configuration approach 
around a distance of $r^*$ or lower, and collide with a 
horizontal head to tail or tail to head approach.  \\

For ${\cal E} = 12.7$~kV/cm in Fig.~\ref{FIG-EPM-2}, 
the energy of the initial CMS is now the highest one.
This is the condition to have for shielding purposes.
For $\theta=0$, 
$E_a$ is attractive and the approach is vertical tail to head
while $E_b$ is repulsive.
As $r$ decreases, the two curves cross at $r=r^*$ and
$E_a$ becomes smaller than $E_b$. The initial CMS
takes over the character of highest energy, $E_b$, as can be seen on the figure,
with a vertical head to tail or tail to head approach.
The corresponding adiabatic energy curve is then $E_+$.
The adiabatic state $\big| + \big\rangle$ in Eq.~\eqref{ESpm} 
is a quantum mechanical linear combination of the two approaches
that depends on $r$.
At large $r$, 
the picture is that of two molecules starting to collide with a vertical tail to head approach.
As the molecule approach, they interlock themselves 
and change of configuration to a vertical head to tail or tail to head approach.
In other words, for the $\theta=0$ approach, 
one of the initial rotational state $n=1$ changes to $n=0$
(qualitatively like a dipole flip from $\downarrow$ to $\uparrow$)
and the other initial rotational state $n=1$
changes to $n=2$
(qualitatively like a dipole $\downarrow$ remaining $\downarrow$).
In that way, the molecules remain in the repulsive curve of highest energy,
where the shield is preserved and effective.
For $\theta=\pi/2$, $E_a$ is repulsive and $E_b$ is attractive.
As $E_a$ remains bigger than $E_b$,
the initial CMS when following adiabatically the curve $E_+$,
will collide with an overall horizontal tail to tail approach,
corresponding to an overall repulsive interaction.
So depending on the angular approach, the initial rotational
states will change and adapt ($\theta=0$) or remain unchanged ($\theta=\pi/2$)
so that to remain on the repulsive adiabatic curve $E_+$.
This microscopic physical interpretation of the configuration approaches
explains how and why the molecules remain shielded
through the collision above the field of the F{\"o}rster resonance
for any angular configurations. \\

To complete the two-level model, we add to the above eigenvalues 
in Eq.~\eqref{Epm} the overall electronic van der Waals interaction of Eq.~\eqref{VvdW}
and the usual centrifugal term in a partial wave expansion 
(see \cite{Wang_NJP_17_035015_2015} for example), so that we can define two
energy surface functions 
\begin{equation}\label{Vpm}
V_{\pm}(r, \theta) = E_{\pm}(r, \theta) - \frac{C_6^{el}}{r^6} 
+ \frac{\hbar^2 l(l+1)}{2 \mu r^2}
\end{equation}
with $\mu$ being the reduced mass of the molecule-molecule system.
\\

\begin{figure*}[th!]
\begin{center}
\includegraphics*[width=12.4cm,trim=0.8cm 2.9cm 1cm 2cm]{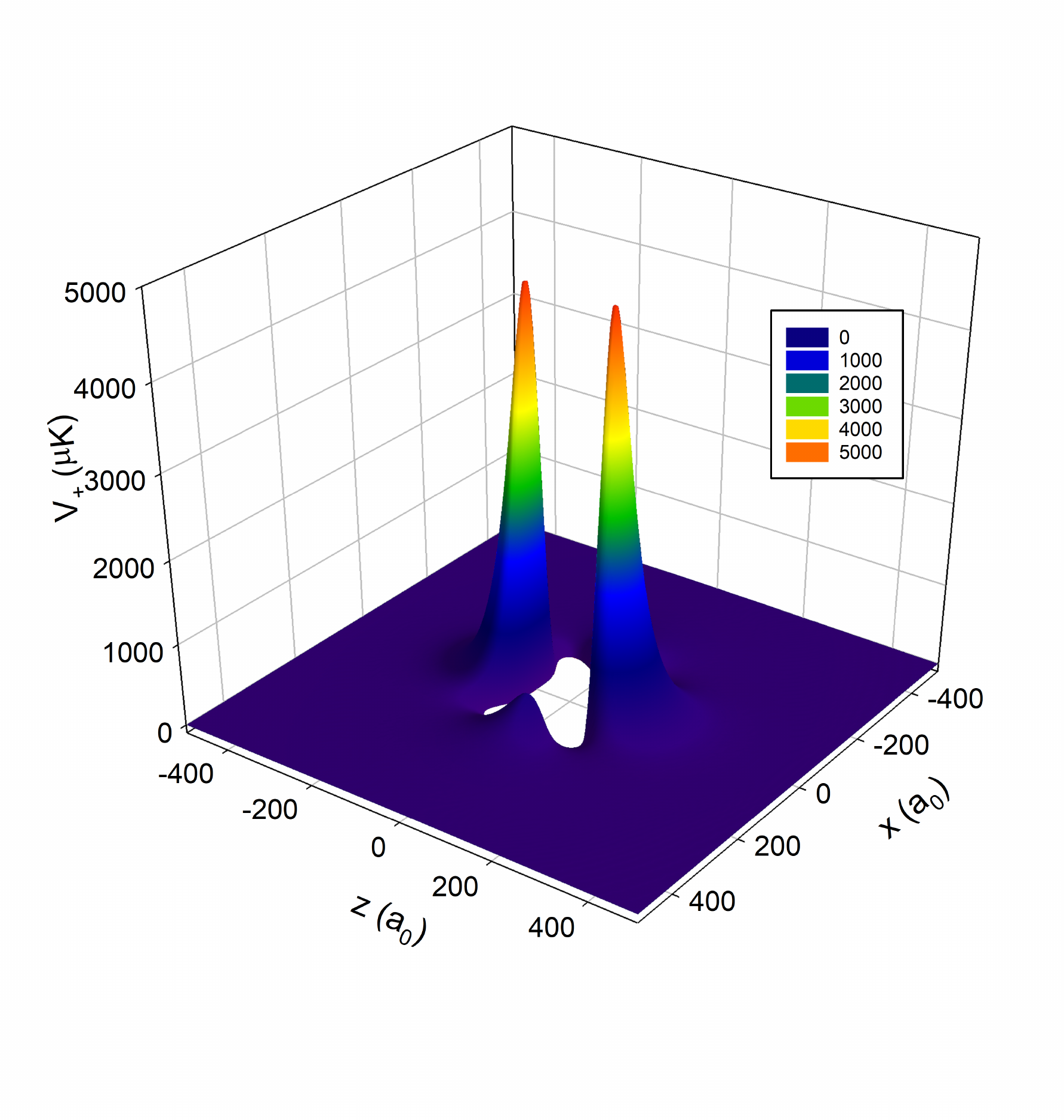} \\
\includegraphics*[width=12.4cm,trim=0cm 2.5cm 1cm 2cm]{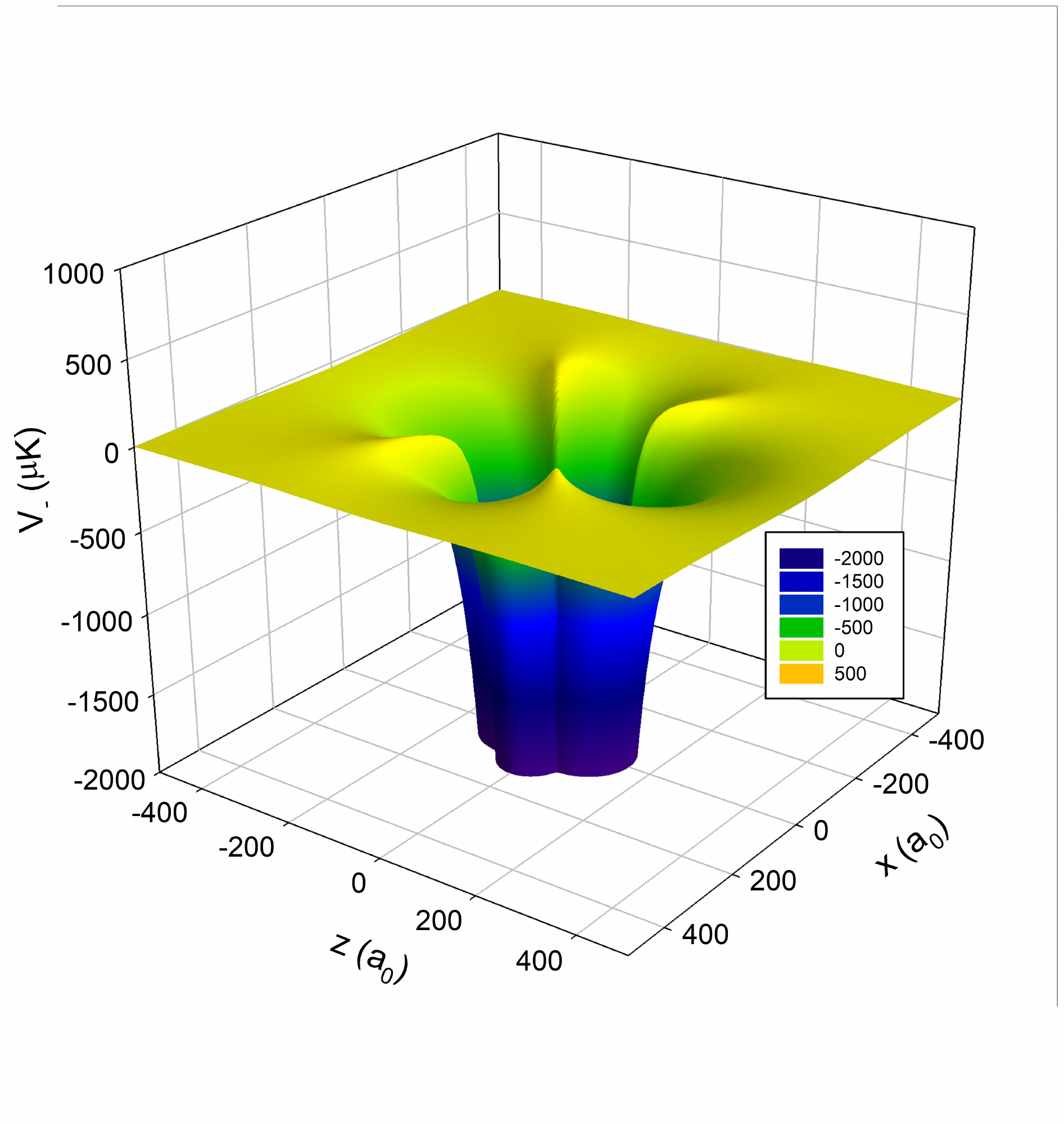} 
\caption{Energy surfaces $V_+$ (top panel) and $V_-$ (bottom panel) 
as a function of the cartesian coordinates $x=r\sin\theta$ and $z=r\cos\theta$. 
They are calculated for fermionic $^{40}$K$^{87}$Rb molecules at respectively 
${\cal E}=12.7$~kV/cm and ${\cal E}=12.5$~kV/cm.
They include the effective dipole-dipole interaction surfaces $E_\pm$, the electronic 
van der Waals interaction and the $l=1$ centrifugial barrier.}
\label{FIG-Vplusminus}
\end{center}
\end{figure*}

In Fig.~\ref{FIG-Vplusminus}, we plot 
two examples of the surfaces defined in Eq.~\eqref{Vpm}
as a function of the cartesian coordinates 
\begin{align*}
x &= r\sin\theta & z &= r\cos\theta.
\end{align*}
The line $x=0$ corresponds to the angle $\theta = 0$
in Eq.~\eqref{Vdd-classic} and to a vertical approach of the dipoles
when $z$ varies.
The line $z=0$ corresponds to the angle $\theta = \pi/2$ 
and to a horizontal approach of the dipoles when $x$ varies.
We took fermionic $^{40}$K$^{87}$Rb molecules as example
so that $l=1$ in Eq.~\eqref{Vpm}.
We present the surfaces $V_+(x,z)$ (top panel) at ${\cal E}=12.7$~kV/cm, 
and $V_-(x,z)$ (bottom panel) at ${\cal E}=12.5$~kV/cm. 
In each case, this is the surface followed adiabatically by the initial CMS. \\

At ${\cal E}=12.5$~kV/cm, one can see that $V_-$ is globally attractive.
This is due to the fact that $E_-$ 
in Eq.~\eqref{Epm} is also globally attractive for the overall configurations.
Only around $\theta\approx \ang{55}$, where $V_{dd}=0$ in Eq.~\eqref{Vdd-classic}, 
the centrifugal and van der Waals terms dominate
and a (small) centrifugal barrier can be seen in the surface.
The overall surface $V_-$ shows that nothing prevents the two 
colliding molecules to come close to each other and to be prone of collisional losses,
in this case, chemical reactivity for KRb molecules. \\

In contrast, at ${\cal E}=12.7$~kV/cm, one can see that $V_+$ is globally repulsive.
We observe four strong peaks which are signatures of the dipole-dipole interaction.
The smaller peaks along $z=0$ are coming from the $\theta=\pi/2$ approach
whereas the two stronger peaks along $x=0$ are coming from the $\theta=0$ one. 
The origin of these peaks comes from the fact that 
$E_+$ in Eq.~\eqref{Epm} is repulsive, as illustrated
in Fig.~\ref{FIG-EPM-2} for $\theta=0,\pi/2$
and discussed above,
where the molecules adapt/arrange their rotational structure
so that the shielding is always effective
for any angular configurations.
We notice that the difference in the height of the peaks between 
$\theta=0$ and $\theta=\pi/2$ is the result of the diagonalization
of the matrix in Eq.~\eqref{2by2mat} and obtention of the eigenvalue $E_+$. 
This difference is explained by the fact that $W$ is proportional to $|\left(1-3 \cos ^{2} \theta\right)|$ which is two times stronger for $\theta=0$ than for $\theta=\pi/2$.
Eventually at lower distance, the van der Waals term in Eq.~\eqref{Vpm}
will be more attractive than $E_+$, explaining 
why these peaks are of finite height and not indefinitely repulsive. 
These surfaces provide an intuitive 
three-dimensional picture of the dynamics around the F{\"o}rster resonance 
and the shielding process. \\

Note that for the case of bosons with $l=0$, there is no centrifugal term in Eq.~\eqref{Vpm}.
The same behavior shown in Fig.~\ref{FIG-Vplusminus} will be observed for both $V_+$ and $V_-$, except that there is no centrifugal barrier in the surface $V_-$, in contrast with what we see
for fermions and $l=1$. But the four peaks will still be present as in $V_+$ so that the shielding survives also for bosons.

%\nopagebreak[0]
%\newpage
\afterpage{\clearpage}
%\clearpage
%\newpage
\onecolumngrid
\twocolumngrid
%\afterpage
%\afterpage{\clearpage}
\clearpage
%\newpage

\subsection{Fourth simplification: Average over the lowest orbital angular momentum}

To complete the model and to compute the values of the observables such as
cross sections or rate coefficients, 
one has to average the surfaces obtained in Eq.~\eqref{Vpm} over the angle $\theta$.
Then, one gets an effective potential as a function of $r$ 
for the initial radial motion of the collision.
When ultracold dynamics occur in the presence of an electric field,
this average is usually perfomed between several partial waves,
described by spherical harmonics functions. 
Here, we will instead
assume that the overall dynamics is entirely described 
by the first and lowest partial wave, namely $l=0$ for indistinguishable bosons
of $l=1$ for indistinguishable fermions.
This is following the previous consideration that the dynamical process comes mainly from the
dipolar interaction at the F{\"o}rster resonance, mainly described by the surfaces $V_+$ and $V_-$ and that the one coming from the orbital motion acts only secondly. \\

The spherical harmonics that describes the partial waves can be separated in a 
polar and azimuthal angular part~\cite{Brandsen_Joachain_Book_2003}
\begin{eqnarray}
{Y}_l^{m_l}(\theta,\phi) = \Theta_l^{m_l}(\theta) \, \Phi_{m_l}(\phi)
\end{eqnarray} 
where for $m_l \ge 0$
\begin{eqnarray}
\Theta_l^{m_l}(\theta)  =  (-1)^{m_l}  \, \sqrt{\frac{2l+1}{2}}  \, \sqrt{\frac{(l-m_l)!}{(l+m_l)!}} \, P_l^{m_l}(\cos\theta) , \nonumber \\
\end{eqnarray} 
for $m_l < 0$
\begin{eqnarray}
\Theta_l^{m_l}(\theta)  = (-1)^{|m_l|}  \, \Theta_l^{|m_l|}(\theta),
\end{eqnarray}
and 
\begin{eqnarray}
\Phi_{m_l}(\phi) = \frac{1}{\sqrt{2 \pi}} \, e^{i \, m_l \, \phi} .
\end{eqnarray}
The functions $\Theta$ and $\Phi$ are normalized 
over the ranges $\theta=[0-\pi]$ and $\phi=[0-2\pi]$. 
For each $r$, the surfaces in Eq.~\eqref{Vpm} are averaged 
over ${Y}_l^{m_l}(\theta,\phi)$ to get an effective potential
for the initial radial motion
\begin{multline}
 \big\langle V_\pm(r) \big\rangle = \big\langle l, m_l \big| V_\pm(r,\theta) \big| l', m_l' \big\rangle \\
 = \int_0^{\pi} \, \int_0^{2 \pi} \ 
 [{Y}_l^{m_l}(\theta,\phi)]^* \, V_\pm(r,\theta) \, {Y}_{l'}^{m_l'}(\theta,\phi) \sin{\theta} \, d\theta \, d\phi.
\end{multline}
Since the potential is independent of $\phi$, then $m_l=m_l'$. 
Therefore for a given $m_l$, the above integral reduces to
\begin{eqnarray}
\label{Eq:1D-quadrature-2B}
 \big\langle V_\pm(r) \big\rangle  = \int_0^{\pi} [\Theta_l^{m_l}(\theta)]^* \, V_\pm(r,\theta) \, \Theta_l^{m_l}(\theta) \sin{\theta} \, d\theta.
\end{eqnarray}
Namely for indistinguishable bosons and $l=0$, we have
\begin{eqnarray}\label{Eq:Adiabatic_curve_l=0}
 \big\langle V_\pm(r)\big\rangle = \int_0^{\pi} \frac{1}{2} \, V_\pm(r,\theta) \, \sin{\theta} \, d\theta.
\end{eqnarray}
For indistinguishable fermions, we have for $l=1, m_l=0$
\begin{eqnarray}\label{Eq:Adiabatic_curve_l=1_ml0}
 \big\langle V_\pm(r) \big\rangle = \int \frac{3}{2} \, V_\pm(r,\theta) \, \cos^2{\theta} \, \sin{\theta} \, d\theta
\end{eqnarray}
and for $l=1, m_l=+1$ or $m_l=-1$
\begin{eqnarray}\label{Eq:Adiabatic_curve_l=1_mlnot0}
 \big\langle V_\pm(r) \big\rangle = \int \frac{3}{4} \, V_\pm(r,\theta) \, \sin^2\theta \, \sin{\theta} \, d\theta.
\end{eqnarray}
\\

\begin{figure}[t]
\begin{center}
\includegraphics*[width=8cm, trim=0cm 0cm 0cm 0cm]{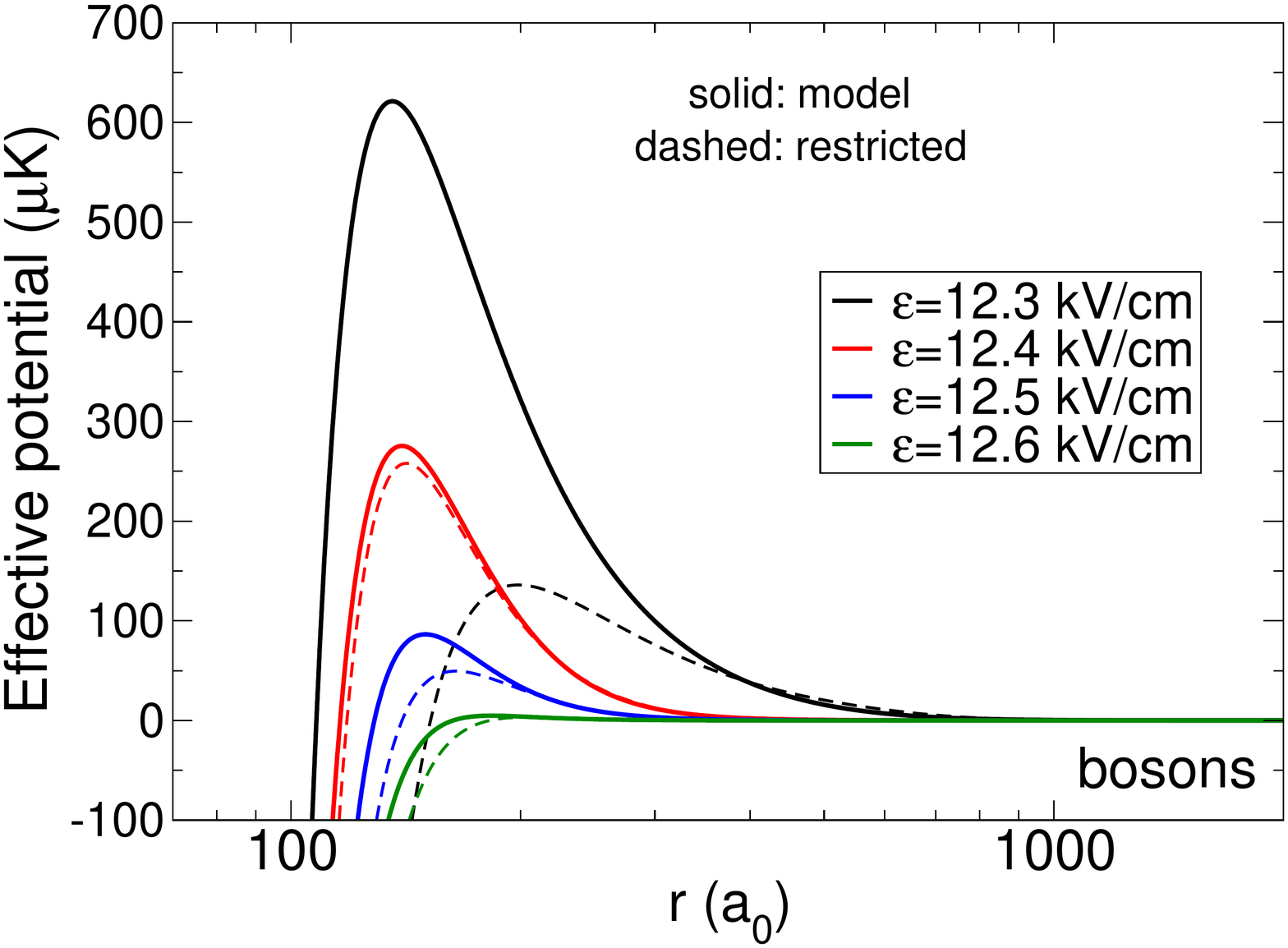} \\
\includegraphics*[width=8cm, trim=0cm 0cm 0cm 0cm]{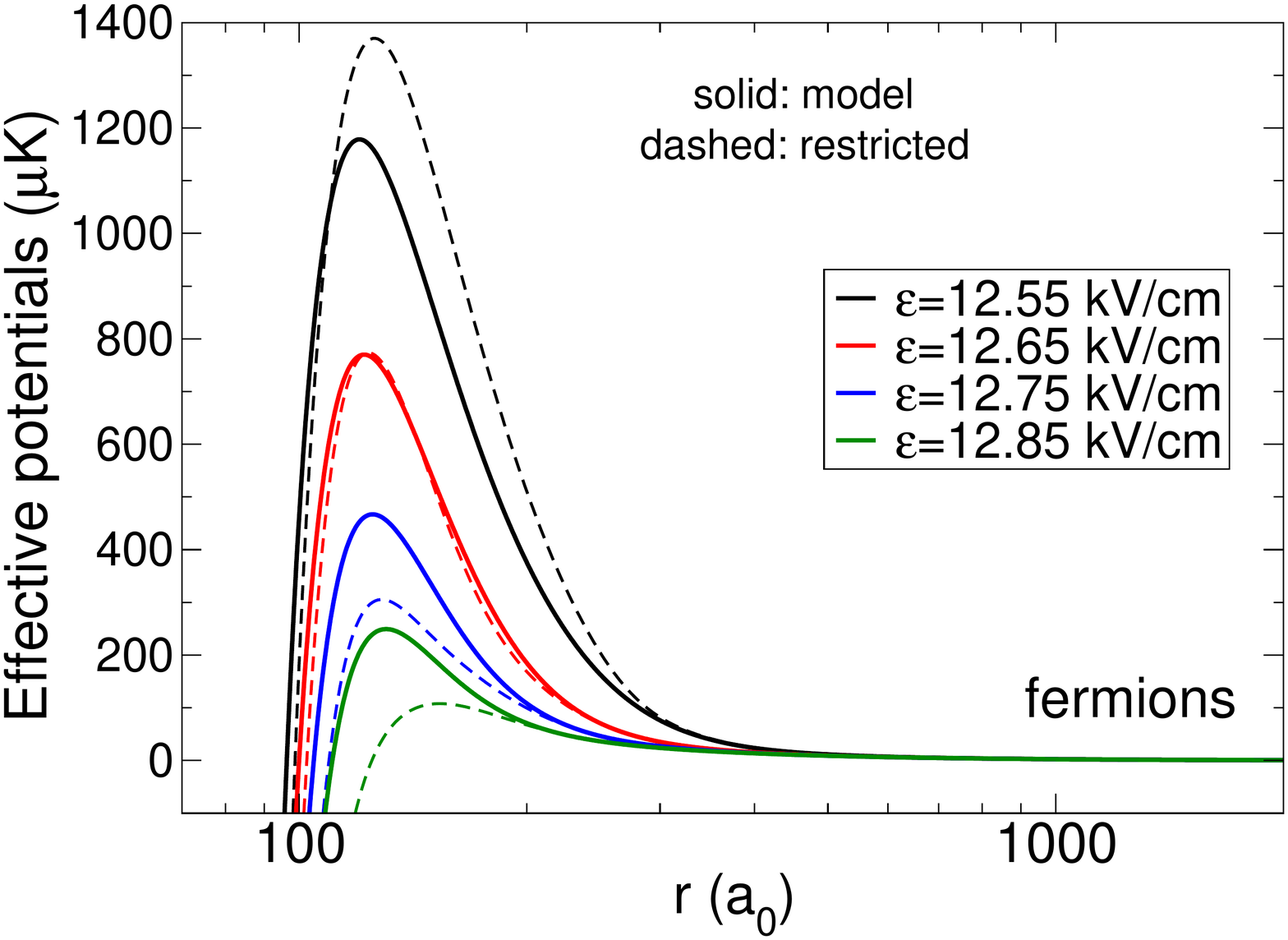} \\
\caption{Effective potentials as a function of $r$ for different electric fields, for the $M = 0$ component. Solid lines: model, dashed lines: restricted calculation. 
Top panel:  bosonic $^{41}$K$^{87}$Rb, bottom panel: fermionic $^{40}$K$^{87}$Rb.
As an indication, the collision energy is on the order of hundreds of nanokelvins while the height of the barriers is on the order of hundreds of microkelvins.}
\label{FIG-ADIAB2}
\end{center}
\end{figure}

We show in Fig.~\ref{FIG-ADIAB2} the effective potential energy curves 
obtained with one channel (solid lines) 
using Eq.~\eqref{Eq:Adiabatic_curve_l=0} 
for bosonic $^{41}$K$^{87}$Rb (top panel)
and with Eq.~\eqref{Eq:Adiabatic_curve_l=1_ml0}
for fermionic $^{40}$K$^{87}$Rb (bottom panel).
We compare these curves with the same effective potentials but obtained 
from Eq.~\eqref{Vdd-full} when $m_{j_1}=m_{j_2}=0$ (dashed lines),
which correspond to the restricted calculation, already mentioned earlier.
The colors correspond to different applied electric fields ${\cal E} > {\cal E}^*$.
Qualitativelly for all cases, a long-range potential barrier is obtained 
showing that the physics of the shielding mechanism is well described within our model.
For bosons, the solid and dashed lines are really similar for ${\cal E}=12.4$~kV/cm.
For the other fields, the curves differ 
and the model does not catch, for example,
the height of the barriers.
However, we are rather interested in the ultracold physics with temperatures and associated
collision energies of hundreds of nanokelvins, compared to the hundreds of microkelvins
energy scale of the barriers.
Therefore, it is more significant to compare 
the long-range part of the curves than their height, where ultralow energies prevail.
Looking at the long-range part, one can see that the solid lines compare
well with the dashed lines. We expect then that the ultralow energy physics
might be well described with the model.
For fermions, we have similar conclusions.
The solid and dashed lines are similar for ${\cal E}=12.65$~kV/cm
while being different for the other fields.
However, the long-range part of the curves remain similar
and as for bosons, one expect a good collisional description 
for fermions at ultralow energies.
Finally, the heights of the barriers after averaging (hundreds of microkelvins)
in Fig.~\ref{FIG-ADIAB2}
are globally smaller than the peaks of $V_+$ without averaging (thousands of microkelvins)
in Fig.~\ref{FIG-Vplusminus}.
This is due to the fact that the $\theta \simeq 0$ region 
where the very high peaks prevail in Fig.~\ref{FIG-Vplusminus} 
is now weighted by the $\sin \theta$ term in the averaging, 
moderating the final height of the effective radial barriers.

\section{Rate coefficients}
\label{section:rates}

From the effective potentials displayed in Fig.~\ref{FIG-ADIAB2} within our model and simplifications presented in Sec.~\eqref{section:model},
we can now perform the ultracold dynamics of those molecules
and obtain the observables measured in
the experiments, such as the rate coefficients.
This is done using the same time-independent quantum formalism 
than in~\cite{Wang_NJP_17_035015_2015}
but now applied on one channel only (the ones shown in Fig.~\ref{FIG-ADIAB2} as solid lines). \\

\begin{figure}[t]
\begin{center}
\includegraphics*[width=8cm, trim=0cm 0cm 0cm 0cm]{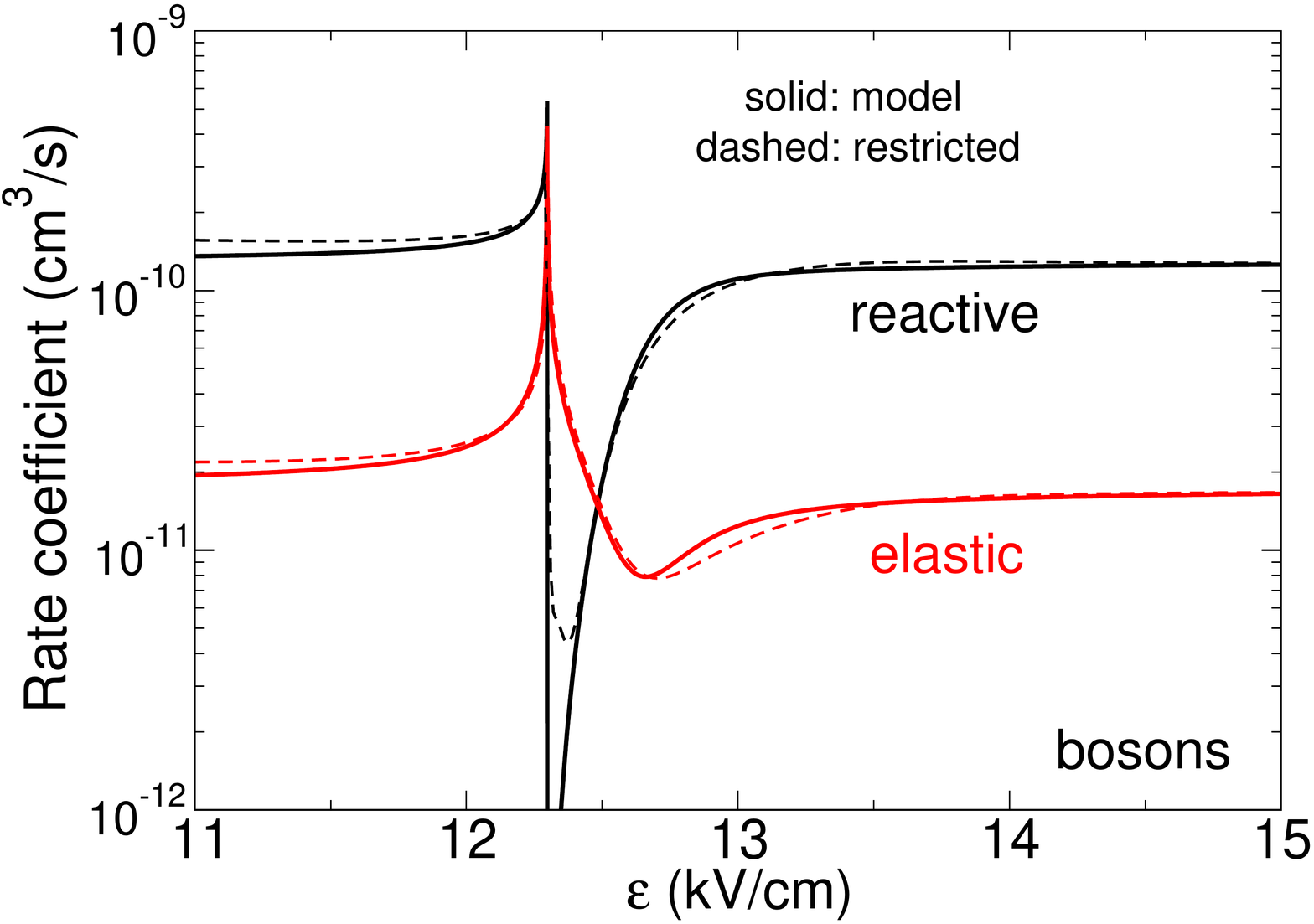} 
\includegraphics*[width=8cm, trim=0cm 0cm 0cm 0cm]{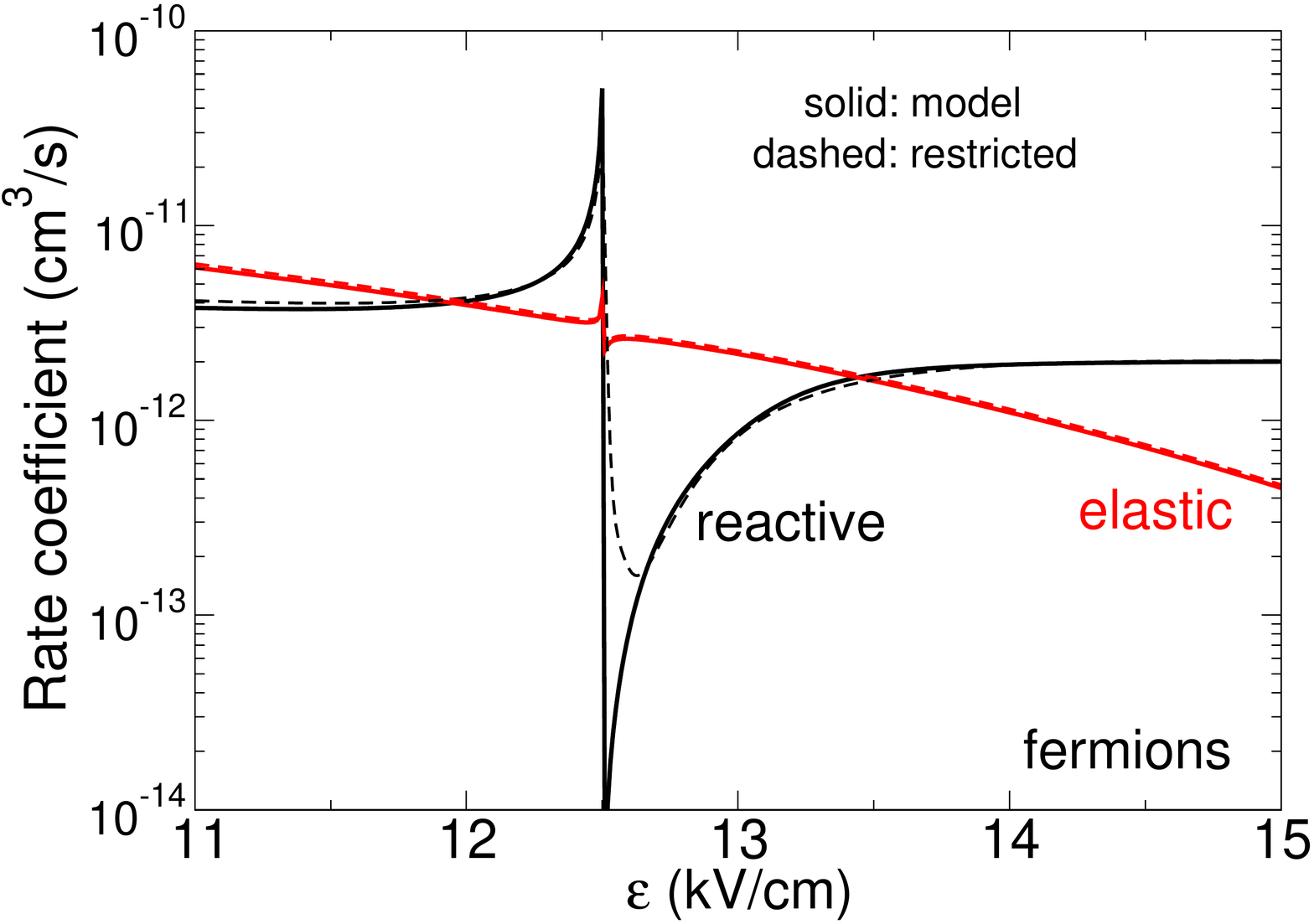} 
\caption{Rate coefficients as a function of the electric field for bosonic $^{41}$K$^{87}$Rb with $M = 0$ (top) and fermionic $^{40}$K$^{87}$Rb with $M = 0,\pm1$ (bottom) at $E_c=500$~nK. Black line: reactive process, red line: elastic process. 
Solid lines: one-channel model, dashed lines: restricted calculation.}
\label{FIG-MODEL-RATE}
\end{center}
\end{figure}

We present in Fig.~\ref{FIG-MODEL-RATE}
the elastic (red) and reactive (black) rate coefficients as a function of the electric field,
obtained within our model (solid lines) consisting on one channel only.
The top (bottom) panel corresponds to bosons for $M=0$ (fermions for $M=0,\pm1$).
In comparison, we provide the same curves but from the restricted 
calculation (dashed lines), involving more channels and partial waves,
and already presented in Fig.~\ref{FIG-FULL} when compared with the full calculation.
One can see that for the range of electric fields around the F{\"o}rster resonance,
the solid lines are very similar to the dashed ones,
showing that the model does a great job to encapsulate the ultracold dynamics of the
molecules, given the simplifications made. 
Only right at the resonance, the reactive processes are underestimated within the model.
We note that the inelastic processes cannot be described by the model as by definition, this is a one-channel model. Therefore, inelastic transitions to another or more channels are absent.
But as seen in Fig.~\ref{FIG-FULL}, the inelastic processes are only dominant right at the resonance where, in any case the model also fails to reproduce the reactive ones.
But as soon as one deviates from the resonance, inelastic processes are negligible
and reactive ones are well described.\\

While the agreement between the model and the restricted calculation is very good, 
the agreement between the model and the full calculation 
can be qualified as semi-quantitative, as the full calculation 
shown in Fig.~\ref{FIG-FULL}
agrees only semi-quantitativelly with the restricted one.
So the discrepancy comes more from the first simplification rather
than the other ones.
When comparing Fig.~\ref{FIG-FULL} and Fig.~\ref{FIG-MODEL-RATE}
slightly above resonance,
the one-channel model gives a slightly overestimated
value of the reactive rate coefficient, about a factor of two, 
while the elastic rate coefficient seems to be well described. 
The model is then sufficient to provide with no numerical efforts a lower limit of the value 
$\gamma = \beta_{el}/\beta_{qu}$, 
the ratio of the elastic over the quenching rate coefficients. 
In current experiments of ultracold molecules, a high value of this ratio is wanted.
This values provides a key information for the experiments. 
It can tell at least if the system is adequate
to drive efficient evaporative cooling and reach for quantum degeneracy,
knowing that this estimated ratio is a lower limit and can be bigger 
(for example about a factor of two for the case of KRb)
when a full calculation is made.
Of course, it does not take too much numerical efforts to use a full calculation 
in this case compared to the model.
But it is important to know first whether such a model for two-body collisions
is valid if one wants to extend it to the study of more particles, for example 
three-body collisions under the same conditions, which might be also of importance for ongoing experiments.

\section{Conclusion}
\label{section:conclusion}

In this paper, we proposed a simple model to describe
the dynamics of ultracold dipolar molecules around a F{\"o}rster resonance,
especially when a collisional shielding takes place.
Based on four assumptions,
this model simplifies the quantum dynamical treatment at its best, yet reproducing
the overal behaviour of the observables.
The model is for example sufficient to predict, with no numerical effort,
whether the elastic processes are more prominent 
than the quenching ones for a given system. It is also
able to provide a lower lmit of the ratio of these processes.
By defining an effective energy surface that encapsulates the relative radial and angular approach of the molecules, it is possible to interpret the shielding
in terms of induced dipole moments. During the shielding, while the molecules start 
in a given initial state, say two dipoles down, they take the quantum character of the resonant state as they approach. They interlock such that a dipole-flip occurs and the interaction remains repulsive. 
Then they never come too close and lossy, quenching collisions do not happen.
When the molecules go away, they turn 
back to two dipoles down from another dipole-flip.
This study and this simple model pave the way for considering more particles, for example 
three-body collisions of molecules, under the same conditions of shielding.

\onecolumngrid
\begin{appendix}
\section{Generalized induced dipole moments when $m_{j_1} = m_{j_2} = 0$}
\label{app_GIDM}

We start from Eq.~\eqref{Vdd-full} and use the first simplification so that
$m_{j_1} = m_{j_2} = 0$. We get
\begin{multline}
\left\langle j_{1}, 0, j_{2}, 0, l, m_{l} \left| V_{dd} \right| j_{1}^{\prime}, 0, j_{2}^{\prime}, 0, l^{\prime}, m_{l}^{\prime} \right\rangle \\
= - \frac{2 d_{1} d_{2}}{4 \pi \epsilon_{0} r^{3}}\ 
 \sqrt{\left(2 j_{1}+1\right)\left(2 j_{1}^{\prime}+1\right)}
\left(\begin{array}{ccc}
j_{1} & 1 & j_{1}^{\prime} \\
0 & 0 & 0
\end{array}\right)^2 \
\sqrt{\left(2 j_{2}+1\right)\left(2 j_{2}^{\prime}+1\right)}
\left(\begin{array}{ccc}
j_{2} & 1 & j_{2}^{\prime} \\
0 & 0 & 0
\end{array}\right)^2 \\
\times (-1)^{m_{l}} \,  \sqrt{(2 l+1)\left(2 l^{\prime}+1\right)}  
\left(\begin{array}{ccc}
l & 2 & l^{\prime} \\
0 & 0 & 0
\end{array}\right)\left(\begin{array}{ccc}
l & 2 & l^{\prime} \\
-m_{l} & 0 & m_{l}^{\prime}
\end{array}\right). \nonumber
\end{multline}
We recall that $d_1, d_2$ are the permanent electric dipole moments of molecules 1, 2.
Noting that
\begin{multline}
\left\langle l, m_{l}\left|\frac{\left(1-3 \cos ^{2} \theta\right)}{4 \pi \epsilon_{0} r^{3}}\right| l^{\prime}, m_{l}^{\prime} \right\rangle =
- \frac{1}{4 \pi \epsilon_{0} r^{3}} \, \frac{4 \sqrt{\pi}}{\sqrt{5}} \, \int_{0}^{2 \pi} \int_{0}^{\pi} [Y_{l}^{m_{l}}(\theta, \phi)]^* \,   Y_{2}^{0}(\theta, \phi) \, Y_{l^{\prime}}^{m_{l}^{\prime}}(\theta, \phi) \, \sin{\theta} \, d\theta \, d\phi \\
=- \frac{2}{4 \pi \epsilon_{0} r^{3}}  \, (-1)^{m_{l}} \,  \sqrt{(2 l+1)\left(2 l^{\prime}+1\right)}  
\left(\begin{array}{ccc}
l & 2 & l^{\prime} \\
0 & 0 & 0
\end{array}\right)\left(\begin{array}{ccc}
l & 2 & l^{\prime} \\
-m_{l} & 0 & m_{l}^{\prime}
\end{array}\right), \nonumber
\end{multline}
we get
\begin{multline}
\left\langle j_{1}, 0, j_{2}, 0, l, m_{l} \left| V_{dd} \right| j_{1}^{\prime}, 0, j_{2}^{\prime}, 0, l^{\prime}, m_{l}^{\prime} \right\rangle 
= \left\langle l, m_{l}\left|\frac{\left(1-3 \cos ^{2} \theta\right)}{4 \pi \epsilon_{0} r^{3}}\right| l^{\prime}, m_{l}^{\prime} \right\rangle \\
\times d_1 \, \sqrt{\left(2 j_{1}+1\right)\left(2 j_{1}^{\prime}+1\right)}
\left(\begin{array}{ccc}
j_{1} & 1 & j_{1}^{\prime} \\
0 & 0 & 0
\end{array}\right)^2 \times d_2 \,
\sqrt{\left(2 j_{2}+1\right)\left(2 j_{2}^{\prime}+1\right)}
\left(\begin{array}{ccc}
j_{2} & 1 & j_{2}^{\prime} \\
0 & 0 & 0
\end{array}\right)^2.  \nonumber
\end{multline}
The internal structure of the molecules is then separated from the orbital motion.
In the presence of an electric field, we should use the dressed states 
Eq.~\eqref{dressedrot}, namely
\begin{align}
\big| \tilde{j}_1, m_{j_1} \big\rangle &= \sum_{j_1} \big| {j_1}, m_{j_1} \big\rangle \, 
\langle {j_1}, m_{j_1} \big| \tilde{j_1}, m_{j_1} \big\rangle &
\big| \tilde{j}_2, m_{j_2} \big\rangle &= \sum_{j_2} \big| {j_2}, m_{j_2} \big\rangle \, 
\langle {j_2}, m_{j_2} \big| \tilde{j_2}, m_{j_2} \big\rangle,  \nonumber
\end{align}
or if we omit the $m_j$ numbers in the notation
\begin{align}
\big| \tilde{j}_1 \big\rangle &= \sum_{j_1} \big| {j_1} \big\rangle \, 
\langle {j_1} \big| \tilde{j_1} \big\rangle &
\big| \tilde{j}_2 \big\rangle &= \sum_{j_2} \big| {j_2} \big\rangle \, 
\langle {j_2} \big| \tilde{j_2}\big\rangle. \nonumber
\end{align}
The dipole-dipole interaction (omitting the $m_j$ numbers) is
\begin{multline}
\left\langle \tilde{j}_{1}, \tilde{j}_{2}, l, m_{l} \left| V_{dd} \right| \tilde{j}_{1}^{\prime}, \tilde{j}_{2}^{\prime}, l^{\prime}, m_{l}^{\prime} \right\rangle 
= \left\langle l, m_{l}\left|\frac{\left(1-3 \cos ^{2} \theta\right)}{4 \pi \epsilon_{0} r^{3}}\right| l^{\prime}, m_{l}^{\prime} \right\rangle \\
\times d_1 \, \sum_{j_1} \, \sum_{j_1'}  \,
\langle \tilde{j_1} \big| {j_1} \big\rangle \,
\langle {j_1'} \big| \tilde{j_1'} \big\rangle \,
\sqrt{\left(2 j_{1}+1\right)\left(2 j_{1}^{\prime}+1\right)}
\left(\begin{array}{ccc}
j_{1} & 1 & j_{1}^{\prime} \\
0 & 0 & 0
\end{array}\right)^2 \\
\times d_2 \, \sum_{j_2} \, \sum_{j_2'}  \,
\langle \tilde{j_2} \big| {j_2} \big\rangle \,
\langle {j_2'} \big| \tilde{j_2'} \big\rangle \,
\sqrt{\left(2 j_{2}+1\right)\left(2 j_{2}^{\prime}+1\right)}
\left(\begin{array}{ccc}
j_{2} & 1 & j_{2}^{\prime} \\
0 & 0 & 0
\end{array}\right)^2.  \nonumber
\end{multline}
If we define a generalized dipole moment $d^{\tilde{j} \to \tilde{j}'}$
as in Eq.~\eqref{induced_dipole_2B}, we obtain
for a given transition
$(\tilde{j}_1, \tilde{j}_2) \to (\tilde{j}_1', \tilde{j}_2')$ 
\begin{eqnarray}
\left\langle  l, m_{l} \left| V_{dd} \right| l^{\prime}, m_{l}^{\prime} \right\rangle 
= d^{\tilde{j}_1 \to \tilde{j}'_1}  \ d^{\tilde{j}_2 \to \tilde{j}'_2} 
\times
 \left\langle l, m_{l}\left|\frac{\left(1-3 \cos ^{2} \theta\right)}{4 \pi \epsilon_{0} r^{3}}\right| l^{\prime}, m_{l}^{\prime} \right\rangle . \nonumber
\end{eqnarray}
This comes back to consider a classical expression of $V_{dd}$
given in Eq.~\eqref{Vdd-classic} using generalized induced dipole moments, including transition dipole moments responsible for inelastic transitions between dressed rotational states in an electric field.

\end{appendix}
\twocolumngrid

\bibliography{../../../BIBLIOGRAPHY/bibliography.bib}

\end{document}